\documentclass{article}
\usepackage{graphicx}
\usepackage{amsmath}
\usepackage{amssymb}
\begin{document}

\begin{center}
    {\Huge{Stability of Periodic, Traveling-Wave Solutions to the Capillary-Whitham Equation}}\\[10pt]
    {\Large{John D.~Carter}}\\[4pt]
    {\Large{Morgan Rozman}}\\[10pt]
    {\large{\today}}
    \end{center}

\date{\today}

\section*{Abstract}
Recently, the Whitham and capillary-Whitham equations were shown to accurately model the evolution of surface waves on shallow water~\cite{Trillo,bidWh}.  In order to gain a deeper understanding of these equations, we compute periodic, traveling-wave solutions to both and study their stability.  We present plots of a representative sampling of solutions for a range of wavelengths, wave speeds, wave heights, and surface tension values.  Finally, we discuss the role these parameters play in the stability of the solutions.

\section{Introduction}

The dimensionless Korteweg-deVries equation (KdV) including surface tension, 
\begin{equation} 
u_t+u_x+\Big{(}\frac{1}{6}-\frac{T}{2}\Big{)}u_{xxx}+2uu_x=0,
\end{equation}
is an asymptotic approximation to the surface water-wave problem in the small-amplitude, long-wavelength limit.  The variable $u=u(x,t)$ represents dimensionless surface displacement, $t$ represents the dimensionless temporal variable, $x$ represents the dimensionless spatial variable, and $T\ge0$ represents the dimensionless coefficient of surface tension (the inverse of the Bond number).  This equation only accurately reproduces the unidirectional, linear phase velocity of the full water wave problem for a small range of wavenumbers near zero.  In order to address this issue, Whitham~\cite{Whitham,Whithambook} proposed a generalization of KdV that is now known as the Whitham equation for water waves.  In dimensionless form, this equation is given by
\begin{equation}
u_t+\mathcal{K}u_x+2uu_x=0,
\label{Whitham}
\end{equation}
where $\mathcal{K}$ is the Fourier multiplier defined by
the symbol
\begin{equation}
\widehat{\mathcal{K}f}(k)=\sqrt{\big{(}1+T k^2\big{)}\frac{\tanh(k)}{k}}~\hat{f}(k).
\label{K}
\end{equation}
We refer to equation (\ref{Whitham}) with $T=0$ as the Whitham equation and (\ref{Whitham}) with $T>0$ as the capillary-Whitham, or cW, equation.  Equation (\ref{Whitham}) reproduces the unidirectional phase velocity of the water wave problem with $T\ge0$.  In summarizing some of the recent work on these equations, we focus on the results that are most directly related to the work we present below.  Ehrnstr\"om \& Kalisch~\cite{EK} proved the existence of and computed periodic, traveling-wave solutions to the Whitham equation.  Sanford {\emph{et al.}}~\cite{WhithamStability} and Johnson \& Hur~\cite{MatVera} established that large-amplitude, periodic, traveling-wave solutions of the Whitham equation are unstable, while small-amplitude, periodic, traveling-wave solutions are stable if their wavelength is long enough.   Moldabayev {\emph{et al.}}~\cite{Moldabayev} presented a scaling regime in which the Whitham equation can be derived from the water wave problem and compared its dynamics with those from other models including the Euler equations.  Hur~\cite{WhithamBreaking} proved that solutions to the Whitham equation will break provided that the initial condition is sufficiently asymmetric.  Deconinck \& Trichtchenko~\cite{BernardOlga} proved that the unidirectional nature of the Whitham equation causes it to miss some of the instabilities of the Euler equations.  Dinvay {\emph{et al.}}~\cite{Dinvay} extended the work of Moldabayev {\emph{et al.}}~\cite{Moldabayev} to include surface tension and show that the capillary-Whitham equation gives a more accurate reproduction of the free-surface problem than the KdV and Kawahara (fifth-order KdV) equations.  Trillo {\emph{et al.}}~\cite{Trillo} compared Whitham predictions with measurements from laboratory experiments and showed that the Whitham equation provides an accurate model for the evolution of initial waves of depression, especially when nonlinearity plays a significant role.  Finally, Carter~\cite{bidWh} compared predictions with another set of laboratory measurements and showed that the Whitham and capillary-Whitham equations both more accurately model the evolution of long waves of depression than do the KdV and Serre (Green-Naghdi) equations.

The remainder of the paper is outlined as follows.  Section \ref{TWStabSection} describes the solutions we examine, their properties, and the linear stability calculations.  Section \ref{Numerics} contains plots of solutions to the Whitham and capillary-Whitham equations, plots of the corresponding stability spectra, and a discussion of these results.  This section contains the main results of the paper.  Section \ref{Summary} concludes the paper by summarizing our results.

\section{Traveling waves and their stability}
\label{TWStabSection}

We consider periodic, traveling-wave solutions of the form
\begin{equation}
	u(x,t)=f(x-ct),
	\label{TWansatz}
\end{equation}
where $f$ is a smooth, periodic function with period $L$.  Ehrnstr\"om \& Kalisch~\cite{EK} proved that the Whitham equation admits solutions of this form and Remonato \& Kalisch~\cite{RemonatoKalisch} computed a variety of cW solutions of this form.  Substituting (\ref{TWansatz}) into (\ref{Whitham}) and integrating once gives
\begin{equation}
	-cf+\mathcal{K}f+f^2=B,
	\label{TWEquation}
\end{equation}
where $B$ is the constant of integration.  This equation is invariant under the transformation 
\begin{equation}
	f\rightarrow f+\gamma,~~~~~ c\rightarrow c+2\gamma,~~~~~ B\rightarrow B+\gamma(1-c-\gamma).
	\label{relations}
\end{equation}
Therefore, we consider the entire family of solutions of the form given in equation (\ref{TWansatz}) by considering only solutions with zero mean, that is solutions such that
\begin{equation}
	\int_0^Lf(z)~dz=0.
	\label{ZeroMeanDef}
\end{equation}

In order to study the stability of these solutions, we change variables to a moving coordinate frame by introducing the coordinates, $z=x-ct$ and $\tau=t$.  In the moving coordinate frame, the cW equation is given by
\begin{equation}
	u_{\tau}-cu_z+\mathcal{K}u_z+2uu_z=0.
	\label{movingWhitham}
\end{equation}
We consider perturbed solutions of the form
\begin{equation}
	u_{\text{pert}}(z,\tau)=f(z)+\epsilon w(z,\tau)+\mathcal{O}(\epsilon^2),
	\label{pertAnsatz}
\end{equation}
where $f(z)$ is a zero-mean, periodic, traveling-wave solution of the cW equation (i.e.~a stationary solution of (\ref{movingWhitham})), $w(z,\tau)$ is a real-valued function, and $\epsilon$ is a small, positive constant.  Substituting (\ref{pertAnsatz}) into (\ref{movingWhitham}) and linearizing gives
\begin{equation}
	w_\tau-cw_z+\mathcal{K}w_z+2fw_z+2f_zw = 0.
	\label{linearizedWhitham}
\end{equation}
Without loss of generality, assume
\begin{equation}
	w(z,\tau) = W(z)\mbox{e}^{\lambda\tau}+c.c.,
	\label{wForm}
\end{equation}
where $W$ is a complex-valued function, $\lambda$ is a complex constant, and $c.c.$ denotes complex conjugate.  Substituting (\ref{wForm}) into (\ref{linearizedWhitham}) and simplifying gives
\begin{equation}
	(c-2f)W^{\prime}-2f^{\prime}W-\mathcal{K}W^{\prime}=\lambda W,
	\label{evalProb2}
\end{equation}
where prime means derivative with respect to $z$.  In operator form, equation (\ref{evalProb2}) can be written as
\begin{equation}
	\mathcal{L}W=\lambda W,~~~~{\text{where}}~~~\mathcal{L}=(c-2f)\partial_z-2f^{\prime}-\mathcal{K}\partial_z.
	\label{evalProb}
\end{equation}
We are interested in finding the set of $\lambda$ that lead to bounded solutions of (\ref{evalProb}).  In other words, we are interested in finding the spectrum, $\sigma$, of the operator $\mathcal{L}$.  The spectrum determines the stability of the solutions.  If $\sigma(\mathcal{L})$ has no elements with positive real part, then the solution is said to be spectrally stable.  If $\sigma(\mathcal{L})$ has one or more elements with positive real part, then the solution is said to be unstable.  Since the capillary-Whitham equation is Hamiltonian, see Hur \& Pandey~\cite{HurPandey}, $\sigma(\mathcal{L})$ is symmetric under reflections across both the real and imaginary axes.  We use this fact as one check on our numerical results.


\section{Numerical results}
\label{Numerics}

In this section, we present plots of periodic, traveling-wave solutions of the Whitham and capillary-Whitham equations along with their stability spectra.  The solutions were computed using a generalization of the method presented by Ehrnstr\"om \& Kalisch~\cite{EK}.  Following the work of Sanford {\emph{et al.}}~\cite{WhithamStability}, the stability of these solutions was computed by the Fourier-Floquet-Hill method of Deconinck \& Kutz~\cite{DK}.

\subsection{The Whitham equation}

In order to best understand the role surface tension plays in the stability of periodic, traveling-wave solutions to the capillary-Whitham equation, we begin by reviewing results from the Whitham equation (i.e.~the zero surface tension case).  Hur \& Johnson~\cite{MatVera} proved that all small-amplitude Whitham solutions with $k<1.145$ (where $k$ is the wavenumber of the solution) are stable while all small-amplitude solutions with $k>1.145$ are unstable.  Sanford {\emph{et al.}}~\cite{WhithamStability} numerically corroborated this result, presented numerical results that suggest that all large-amplitude Whitham solutions are unstable, and established that $2\pi$-periodic, traveling-wave solutions with "small" wave heights are spectrally stable while those with "large" wave heights are unstable.  

\begin{figure}
	\centering
	\includegraphics[width=12cm]{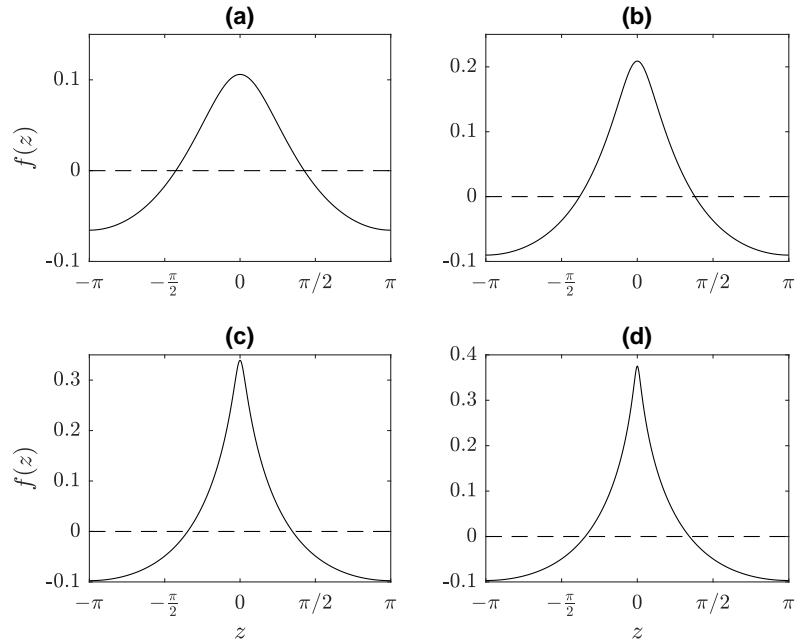}
	\caption{Plots of four moderate wave height, $2\pi$-periodic, zero-mean solutions of the Whitham equation.  The wave speeds and wave heights of these solutions are (a) $c=0.89236$, $H=0.17148$, (b) $c=0.92685$, $H=0.29901$, (c) $c=0.96612$, $H=0.43667$, and (d) $c=0.97249$, $H=0.47203$.  Note that the vertical scale is different in each of the plots.}
	\label{Wsolns}
\end{figure}

\begin{figure}
	\centering
	\includegraphics[width=12cm]{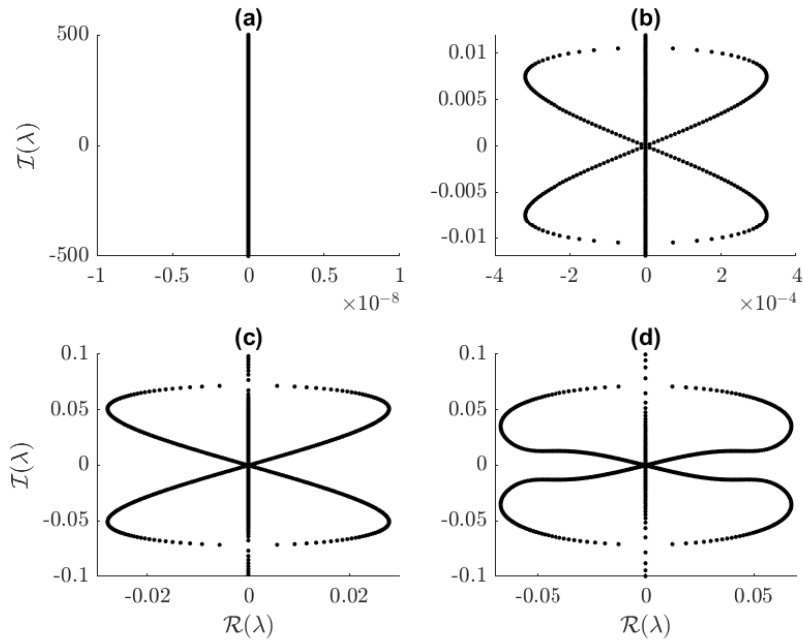}
	\caption{Spectra of the solutions shown in Figure \ref{Wsolns}.  Note that both the horizontal and vertical scales vary from plot to plot.}
	\label{Wsolnsstab}
\end{figure}

Figure \ref{Wsolns} contains plots of four $2\pi$-periodic solutions to the Whitham equation with moderate wave heights.  As the wave height, $H$, of the solution increases, so does the solution's wave speed, $c$.  Figure \ref{Wsolnsstab} contains plots of the stability spectra corresponding to these solutions.  The spectrum of the solution in Figure \ref{Wsolns}(a), see Figure \ref{Wsolnsstab}(a), lies entirely on the imaginary axis and therefore this solution is spectrally stable.  Further simulations (not included) show that all solutions with smaller wave heights (with period $2\pi$) are also spectrally stable.  The spectra corresponding to the other three solutions all include eigenvalues with positive real parts and therefore they are unstable.  As the wave height of the solution increases, the maximum instability growth rate (i.e.~the real part of the eigenvalue with maximal real part) also increases.  All of these spectra include the "figure 8" associated with the modulational (Benjamin-Feir) instability.

\begin{figure}
	\centering
	\includegraphics[width=12cm]{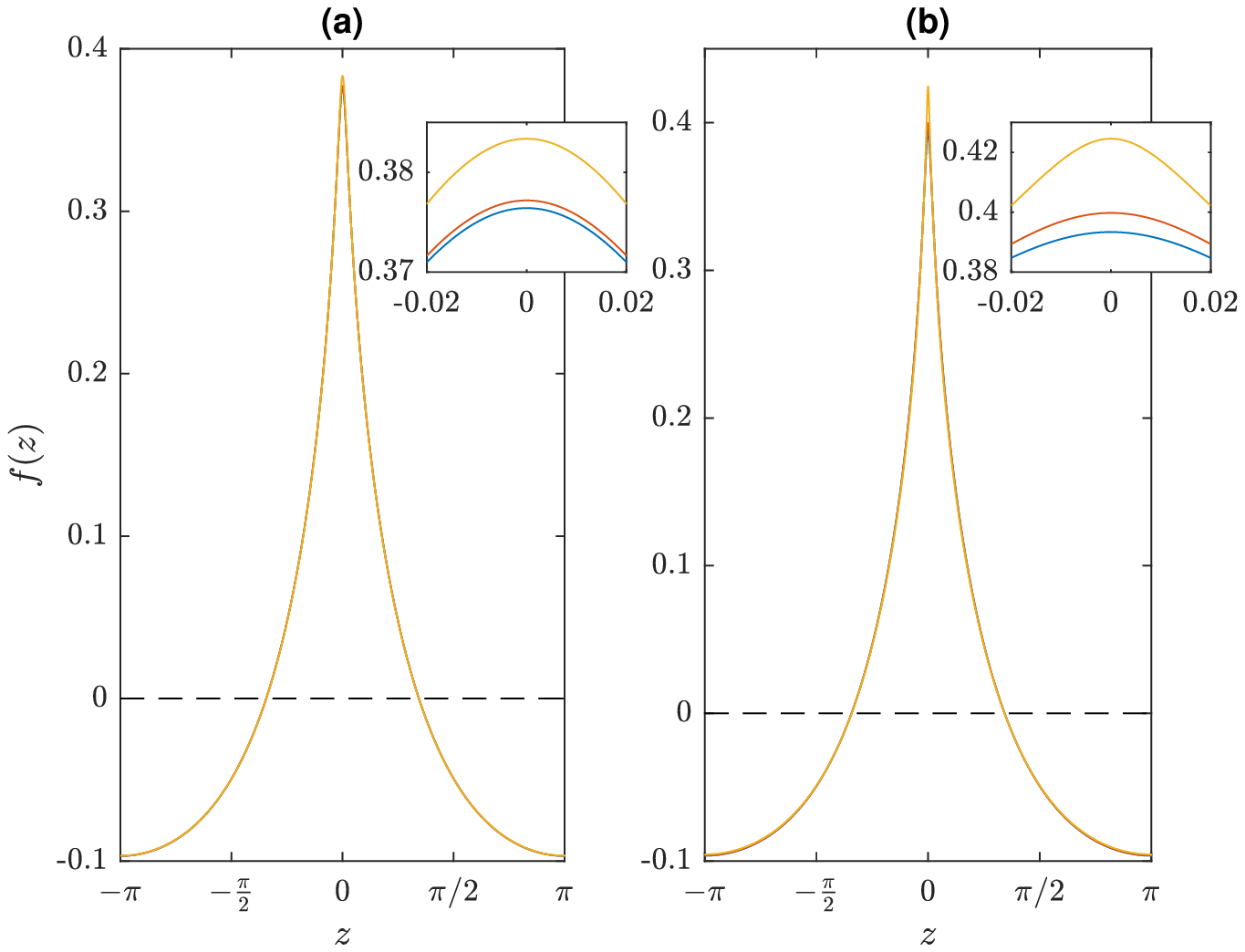}
	\caption{Figures (a) and (b) each contain three plots of large wave height, $2\pi$-periodic, zero-mean solutions of the Whitham equation.  The solutions are very similar and essentially lie on top of one another.  The inset plots are zooms of the intervals surrounding the peaks of the solutions.  The wave speeds and heights of these solutions, in order of increasing speed, are (a) blue $c=0.97266$, $H=0.47330$; orange $c=0.97276$, $H=0.47405$; and yellow $c=0.97351$, $H=0.48007$; and (b) blue $c=0.97451$, $H=0.50058$; orange $c=0.97501$, $H=0.499599$; and yellow $c=0.97596$, $H=0.52013$.}
	\label{WsolnsLarge}
\end{figure}

Whitham~\cite{Whithambook} conjectured that the Whitham equation admits a highest traveling-wave solution and that it is nonsmooth.  Recently, Ehrnstr\"om and Wahl\'en~\cite{EhrnstromWahlen} proved this hypothesis.  Figure \ref{WsolnsLarge} includes plots of six solutions that are somewhat near this highest wave.  The inset plots are zooms of the solutions near their peaks and shows that all of the solutions we consider are smooth.  Note that it is computationally expensive to study solutions near the highest wave due to the number of Fourier modes required to accurately resolve the solutions.  To our knowledge, this is the first time that the stability of solutions of the Whitham equation with wave heights this large have been studied.

\begin{figure}
	\centering
	\includegraphics[width=12cm]{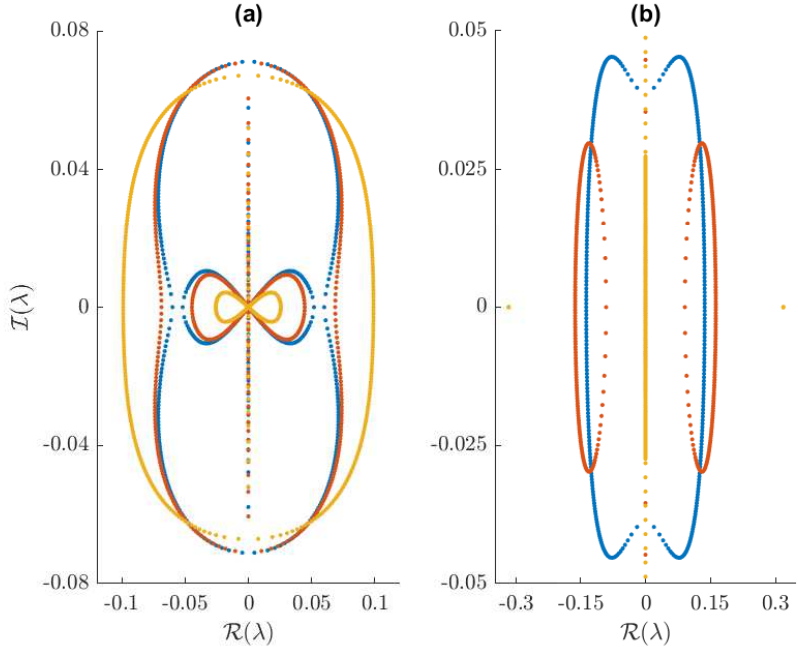}
	\caption{Spectra of the solutions shown in Figure \ref{WsolnsLarge}.}
	\label{WsolnsLargestab}
\end{figure}

Figure \ref{WsolnsLargestab} includes the stability spectra corresponding to the solutions in Figure \ref{WsolnsLarge}.  All six of these solutions are unstable.  As wave height (or wave speed) increases, the maximal instability growth rate increases.  The stability spectra undergo two bifurcations as the wave height increases.  The first bifurcation is shown in Figure \ref{WsolnsLargestab}(a) and the second is shown in Figure \ref{WsolnsLargestab}(b).  The first occurs when the top part of the figure 8 bends down and touches the bottom part.  (See the transition from the spectrum in Figure \ref{Wsolnsstab}(d) to the blue spectrum in Figure \ref{WsolnsLargestab}(a).)  This causes the (vertical) figure 8 to transition into a horizontal figure 8 inside of a vertical "peanut".  (See the orange spectrum in Figure \ref{WsolnsLargestab}(a).)  The second bifurcation occurs when the horizontal figure 8 collapses toward the origin and the peanut pinches off into two ovals centered on the $\mathcal{R}(\lambda)$ axis.  (See Figure \ref{WsolnsLargestab}(b).)  Note that the two yellow "dots" near $\lambda=\pm0.32$ are actually small ovals.  As wave height increases even further, these ovals decrease in diameter and move further away from the $\mathcal{I}(\lambda)$ axis.  Exactly what happens to the stability spectra as the wave height approaches the maximal wave height remains an open question.

\subsection{The capillary-Whitham equation}

In this subsection, we study periodic, traveling-wave, zero-mean solutions of the cW equation and their stability.  Due to the massive number of solutions this equation admits, our study is not meant to be exhaustive.  We present plots of solutions and their stability spectra and end with a discussion that summarizes our observations.  Note that the solutions presented herein cannot be directly compared with those of Remonato \& Kalisch~\cite{RemonatoKalisch} because we required the solutions to have zero mean while they did not.  However, the two sets of solutions are related by equation (\ref{relations}).  

\begin{figure}
	\centering
	\includegraphics[width=12cm]{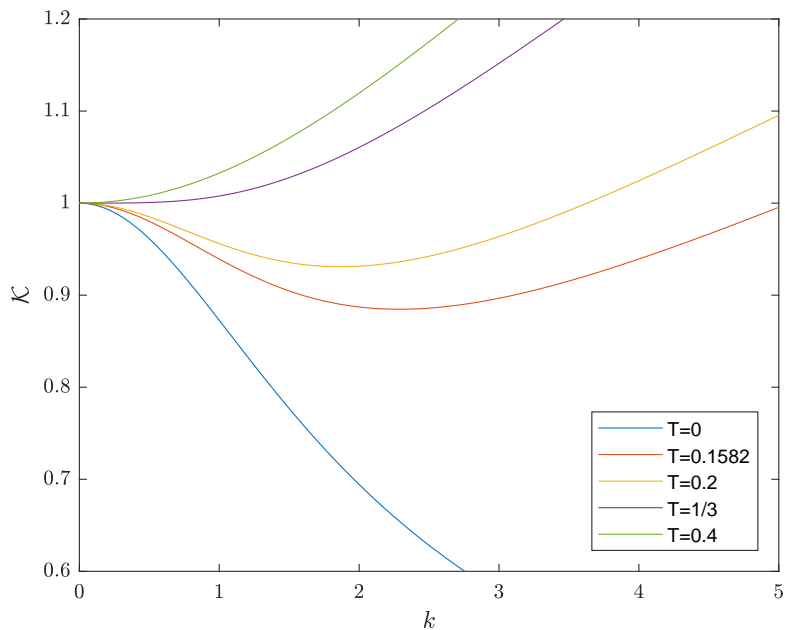}
	\caption{Plots of $\mathcal{K}(k)$ versus $k$ for each of the five values of $T$ examined herein.}
	\label{DispPlots}
\end{figure}

We begin by justifying the values we selected for the capillarity/surface tension parameter, $T$.  The Fourier multiplier $\mathcal{K}$ undergoes a bifurcation at $T=\frac{1}{3}$.  When $T$=0, $\mathcal{K}$ decreases monotonically to zero as the wavenumber of the solution, $k>0$, increases.  When $T\in(0,\frac{1}{3})$, $\mathcal{K}$ achieves a unique local minimum at some wavenumber $k^*\in(0,\infty)$.  When $T>\frac{1}{3}$, $\mathcal{K}$ increases monotonically for all $k>0$ and therefore there is no local minimum.  Because of this behavior, we selected $T=0.2,~\frac{1}{3},$ and $0.4$.  Additionally, we study solutions for $T\approx0.1582$ (see Section \ref{specialT} for details).  Figure \ref{DispPlots} contains plots of $\mathcal{K}$ versus $k$ for each of these $T$ values and demonstrates the bifurcation.

\subsubsection{Surface tension parameter $T=0.2$}

\begin{figure}
	\begin{center}
		\includegraphics[width=12cm]{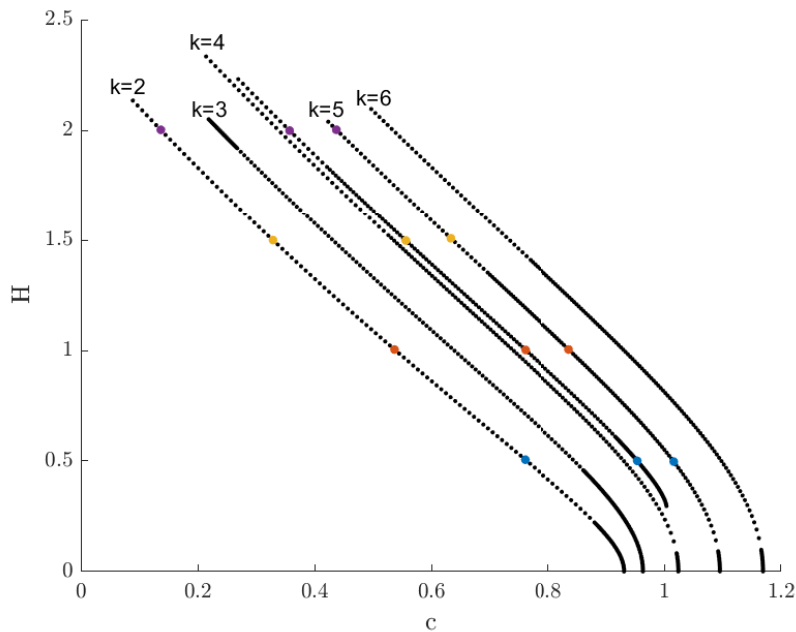}
		\caption{A portion of the wave height versus wave speed bifurcation diagram for the cW equation with $T=0.2$.  The colored dots correspond to solutions that are examined in Figures \ref{fig:twokTwoSolutions}-\ref{fig:twokOneSolutions}.}
		\label{fig:twoBifurcation}
	\end{center}
\end{figure}

When $T=0.2$, we were able to compute solutions with wavenumbers greater than $k\approx1.9$, but were not able to compute solutions with wavenumber less than $k\approx1.9$.  This is interesting because $k^*\approx1.9$.  (Recall that $k^*$ is the location of the local minimum of $\mathcal{K}$ when $T<\frac{1}{3}$.)  Figure \ref{fig:twoBifurcation} includes a portion of the wave height versus wave speed bifurcation diagram for this case.  It includes the bifurcation branches corresponding to the $k=2,\dots, 6$ solutions as well as an additional branch that splits off from the $k=4$ branch when $H>0$.  The colored dots correspond to solutions that are examined in more detail below.  Unlike the Whitham ($T=0$) case, the diagram shows that as the speed of the solutions decreases, the wave height of the solutions increases.  Additionally, it is unclear if any of the branches have upper bounds.

\begin{figure}
	\begin{center}
	\includegraphics[scale=0.4]{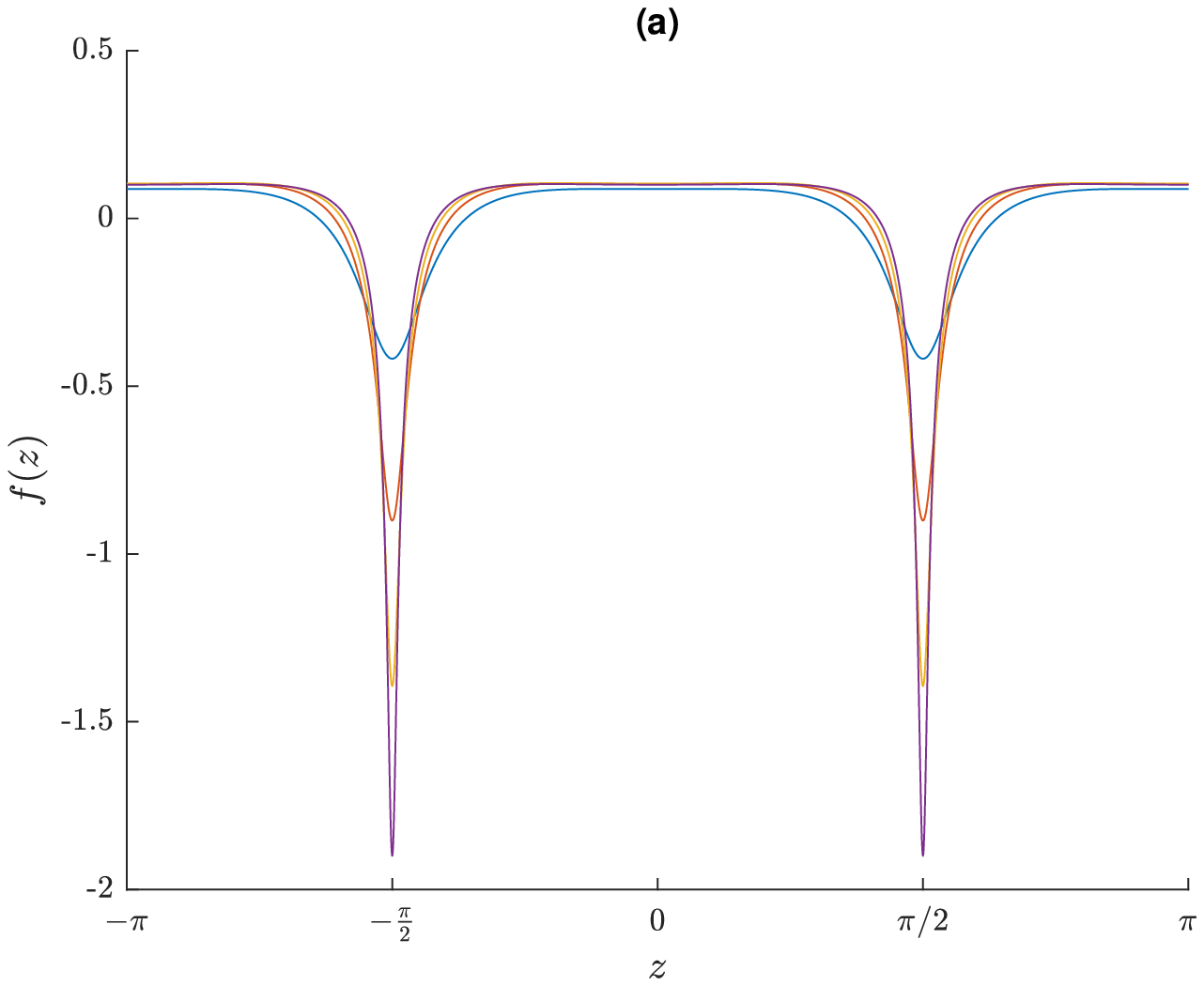}
	\includegraphics[scale=0.4]{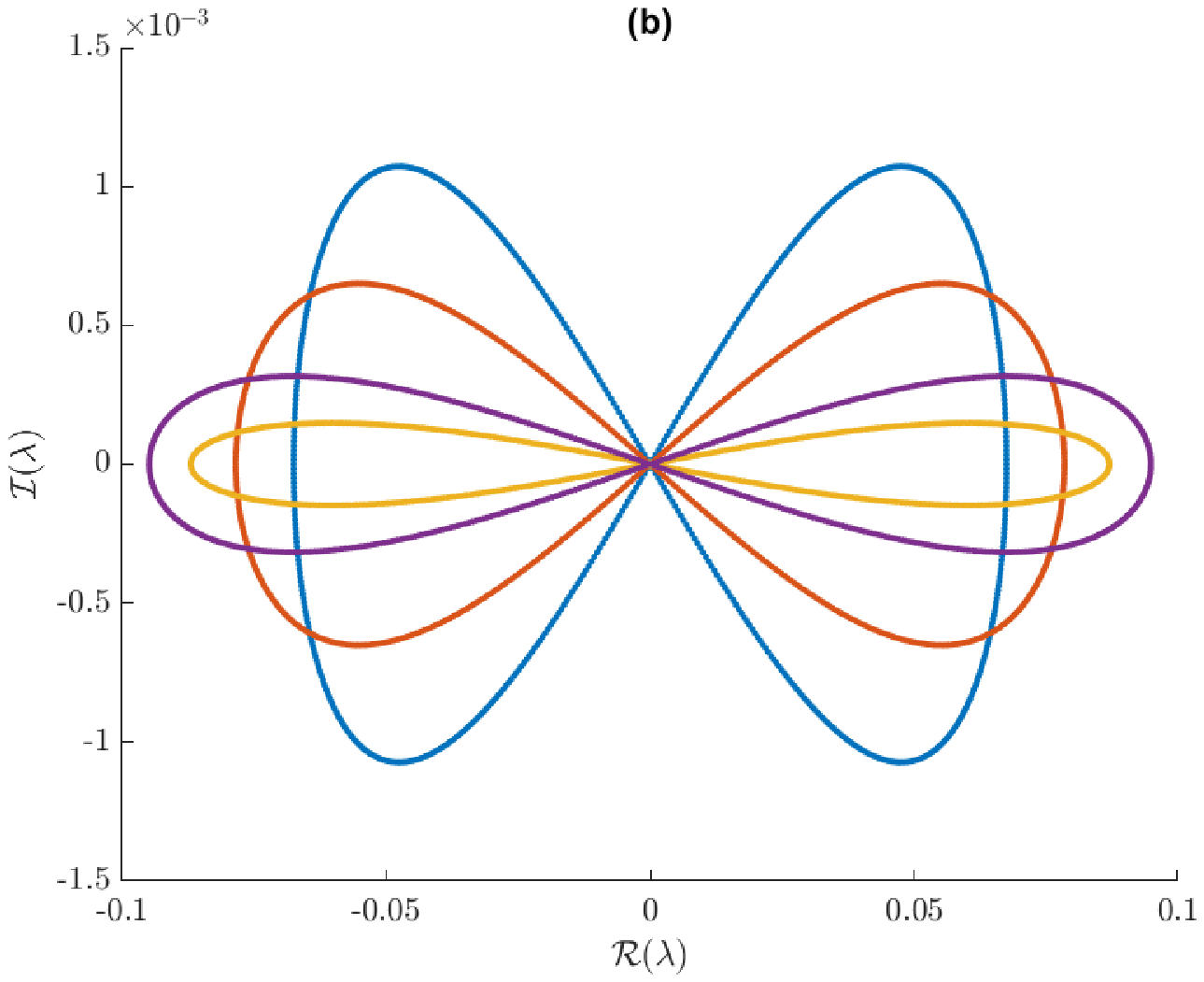}
	\caption{Plots of (a) four representative solutions of the cW equation with $T=0.2$ and $k=2$ and (b) their stability spectra.}
	\label{fig:twokTwoSolutions}
	\end{center}
\end{figure}

Figure \ref{fig:twokTwoSolutions}(a) includes plots of four different $k=2$ (i.e.~period $\pi$), traveling-wave solutions to the cW equation with $T=0.2$.  Unlike solutions of the Whitham equation, these solutions are waves of depression instead of waves of elevation.  As the wave height increases, the solution speed decreases.  Although the bifurcation diagram suggests that a maximal wave height does not exist, we were not able to prove it, numerically or otherwise.  As wave speed decreases, the solutions appear to be approaching the sum of a pair of negative delta functions.  Figure \ref{fig:twokTwoSolutions}(b) contains plots of the corresponding linear stability spectra.  All four of these solutions are unstable.  The figure 8 has switched from being vertical (in the Whitham case) to horizontal (in the cW case).  Just as with solutions to the Whitham equation, the maximal instability growth rate of these solutions increases as their wave heights increase.  Additional numerical simulations (not shown) establish that all small-amplitude, traveling-wave solutions with $k=2$ and $T=0.2$ are unstable.

\begin{figure}
	\begin{center}
	\includegraphics[scale=0.4]{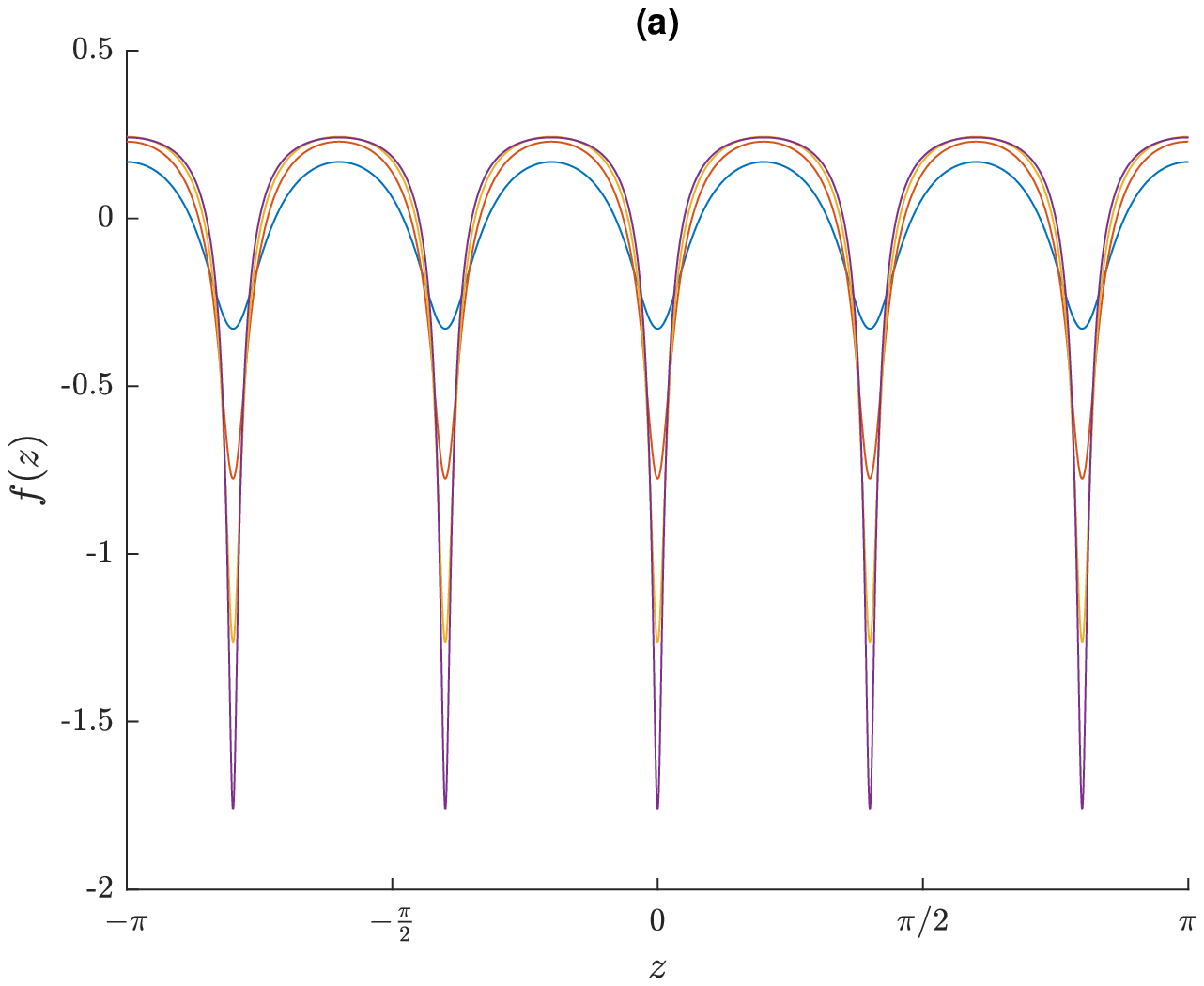}
	\includegraphics[scale=0.4]{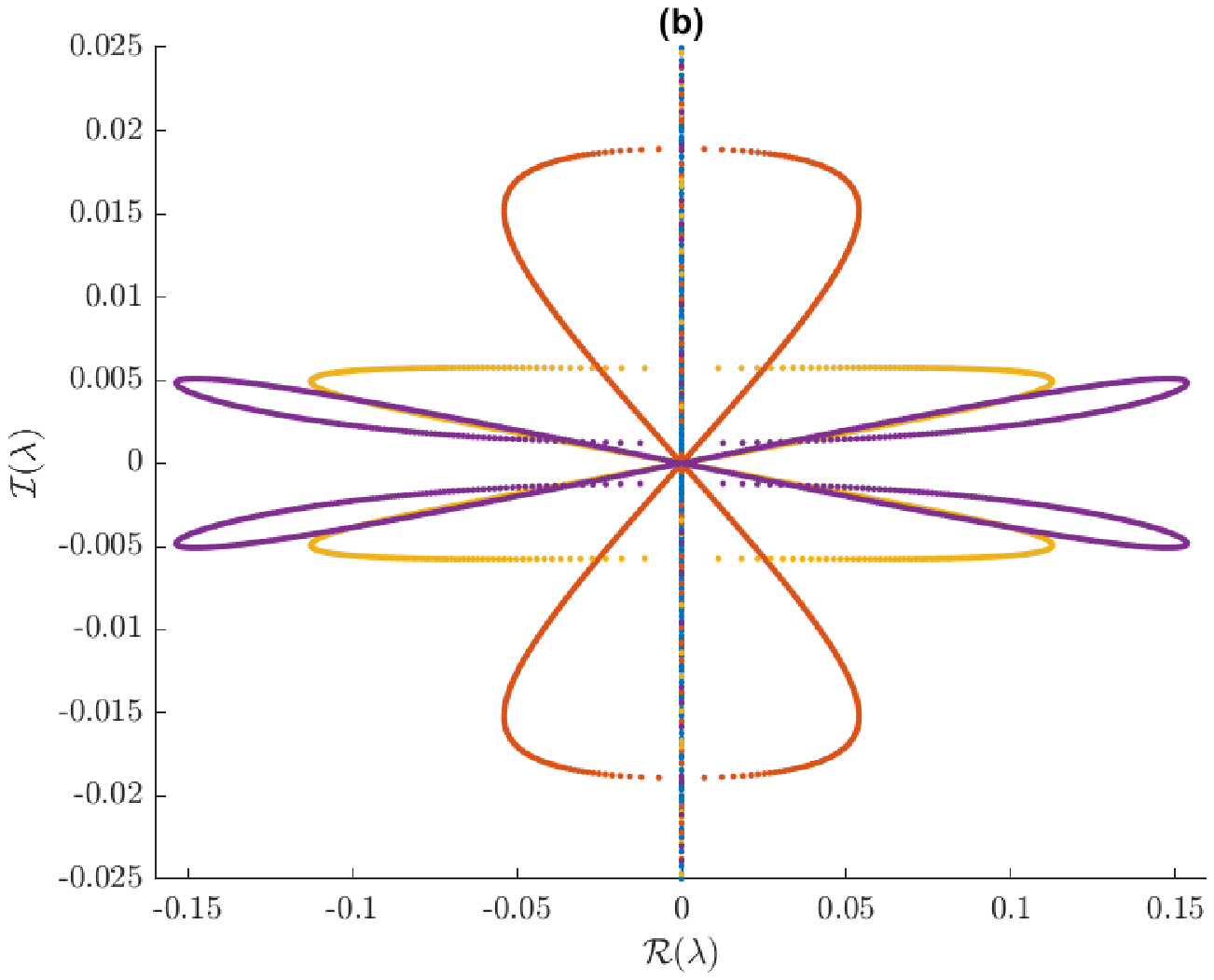}
	\caption{Plots of (a) four representative solutions of the cW equation with $T=0.2$ and $k=5$ and (b) their stability spectra.}
	\label{fig:twokFiveSolutions}
	\end{center}
	\end{figure}

Figure \ref{fig:twokFiveSolutions}(a) includes plots of four $k=5$, traveling-wave solutions to the cW equation with $T=0.2$.  Figure \ref{fig:twokTwoSolutions}(b) shows the corresponding stability spectra.  These four solutions have approximately the same wave heights as the four $k=2$ solutions shown in Figure \ref{fig:twokTwoSolutions}(a).  Other than their period, the $k=5$ solutions are qualitatively similar to those with $k=2$.  The solution with the smallest wave height (blue) is spectrally stable.  This is qualitatively different than what happens in the $T=0$ case where all small-amplitude solutions with $k>1.145$ are unstable.  This suggests that surface tension provides a stabilizing effect to small-amplitude solutions with higher wavenumbers in this case.  The three solutions with larger wave height are unstable and the maximal instability growth rate increases with wave height.  Additional numerical simulations (not shown) establish that solutions with $k\in(2,5)$ have similar properties to the solutions presented in Figures \ref{fig:twokTwoSolutions}-\ref{fig:twokFiveSolutions}.  There exists a $k^\dagger\in(2,5)$ where the small-amplitude solutions switch from being unstable to stable.

\begin{figure}
	\begin{center}
	\includegraphics[scale=0.4]{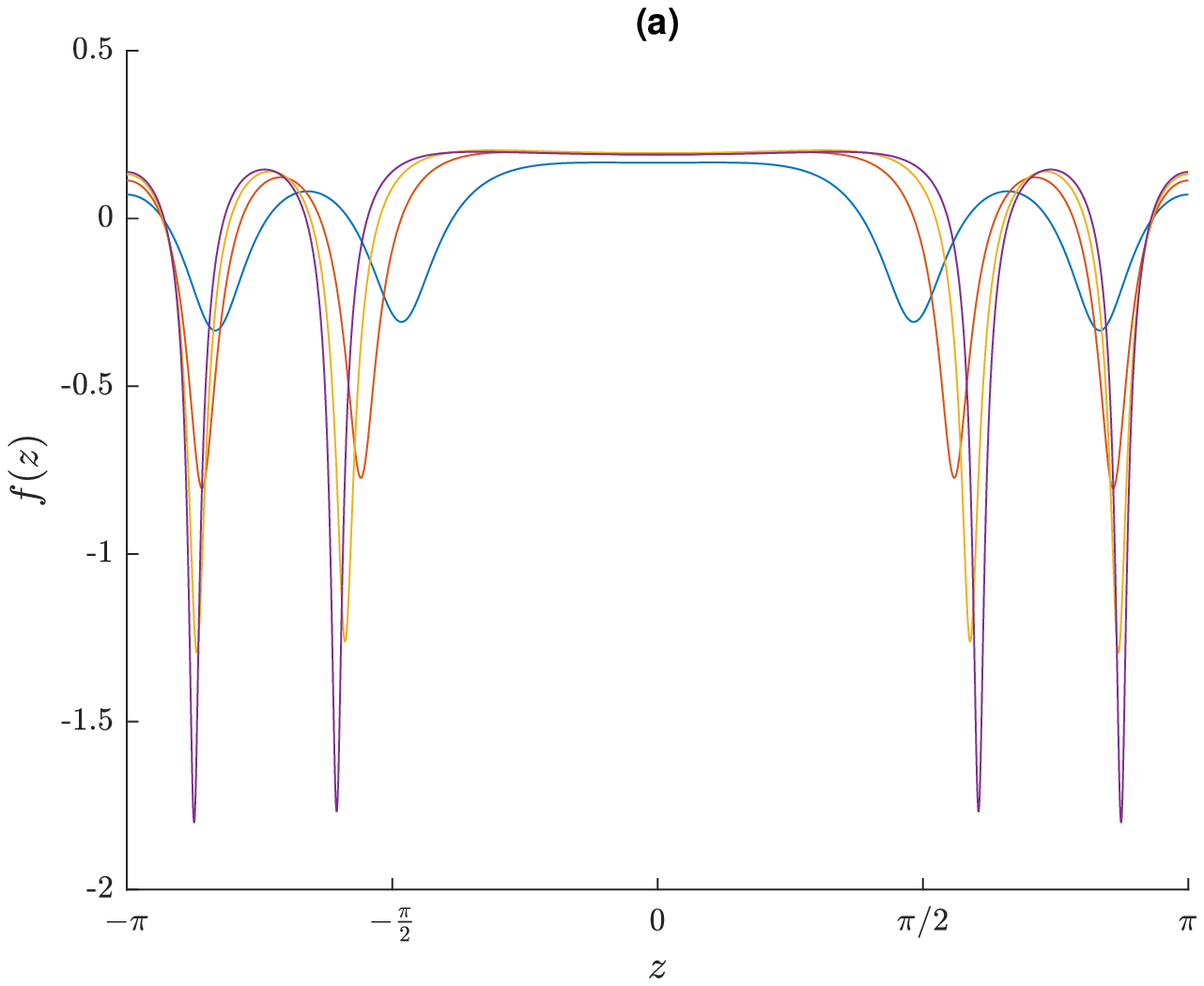}
	\includegraphics[scale=0.4]{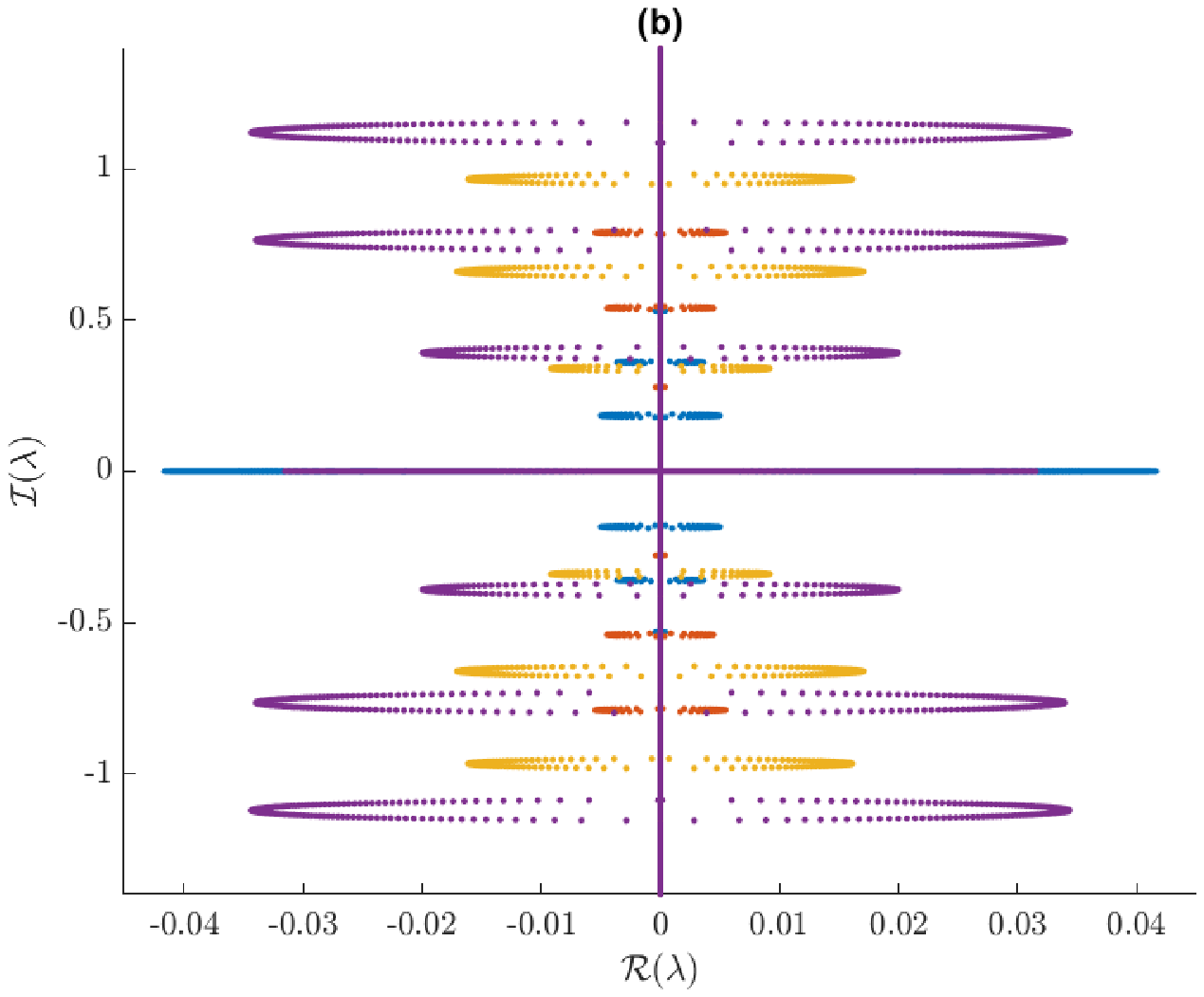}
	\caption{Plots of (a) four representative solutions of the cW equation with $T=0.2$ from the solution branch that does not touch the horizontal axis in Figure \ref{fig:twoBifurcation} and (b) their stability spectra.}
	\label{fig:twokOneSolutions}
	\end{center}
\end{figure}

Figure \ref{fig:twokOneSolutions} contains plots of four representative solutions from the bifurcation branch that splits off from the $k=4$ branch.  These solutions are qualitatively different than the solutions examined above, but have approximately the same wave heights as the solutions in Figures \ref{fig:twokTwoSolutions}(a) and \ref{fig:twokFiveSolutions}(a).  Unsurprisingly, the stability spectra are also qualitatively different than those examined above.  The spectra for each solution includes a horizontal figure 8 centered at the origin.  (In the figure, these appear as horizontal lines along $\mathcal{I}(\lambda)=0$ due to scaling.)  Each spectrum has six additional "bubbles" centered on the $\mathcal{I}(\lambda)$ axis.  (Only four of the blue bubbles are easily visible due to the scaling used.)  Surprisingly, there does not appear to be a simple relationship between wave height and the maximum instability growth rate.  The solution with smallest wave height (the blue solution) has the largest maximum instability growth rate.

\subsubsection{Surface tension parameter $T=\frac{1}{3}$}

\begin{figure}
	\begin{center}
	\includegraphics[width=12cm]{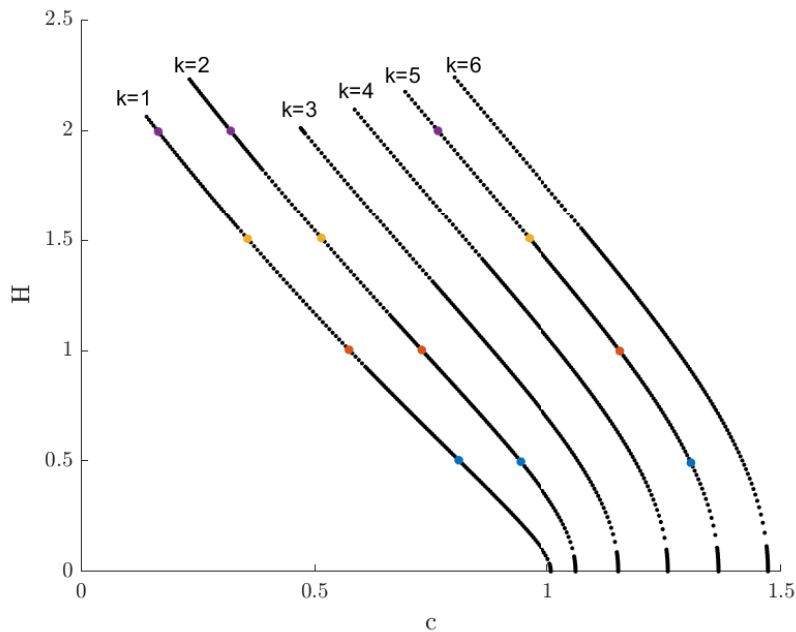}
	\caption{A portion of the bifurcation diagram for the capillary Whitham equation with $T=\frac{1}{3}$.  The colored dots correspond to solutions that are examined in more detail below.}
	\label{fig:thirdBifurcation}
	\end{center}
\end{figure}

Figure \ref{fig:thirdBifurcation} includes a portion of the bifurcation diagram for $T=\frac{1}{3}$.  The colored dots correspond to solutions that are examined in more detail below.  These solutions have approximately the same wave heights as the colored solutions examined in other sections.  The bifurcation diagram shows that as wave speed decreases, wave height increases for all branches (that we examined).

\begin{figure}
	\begin{center}
	\includegraphics[scale=0.4]{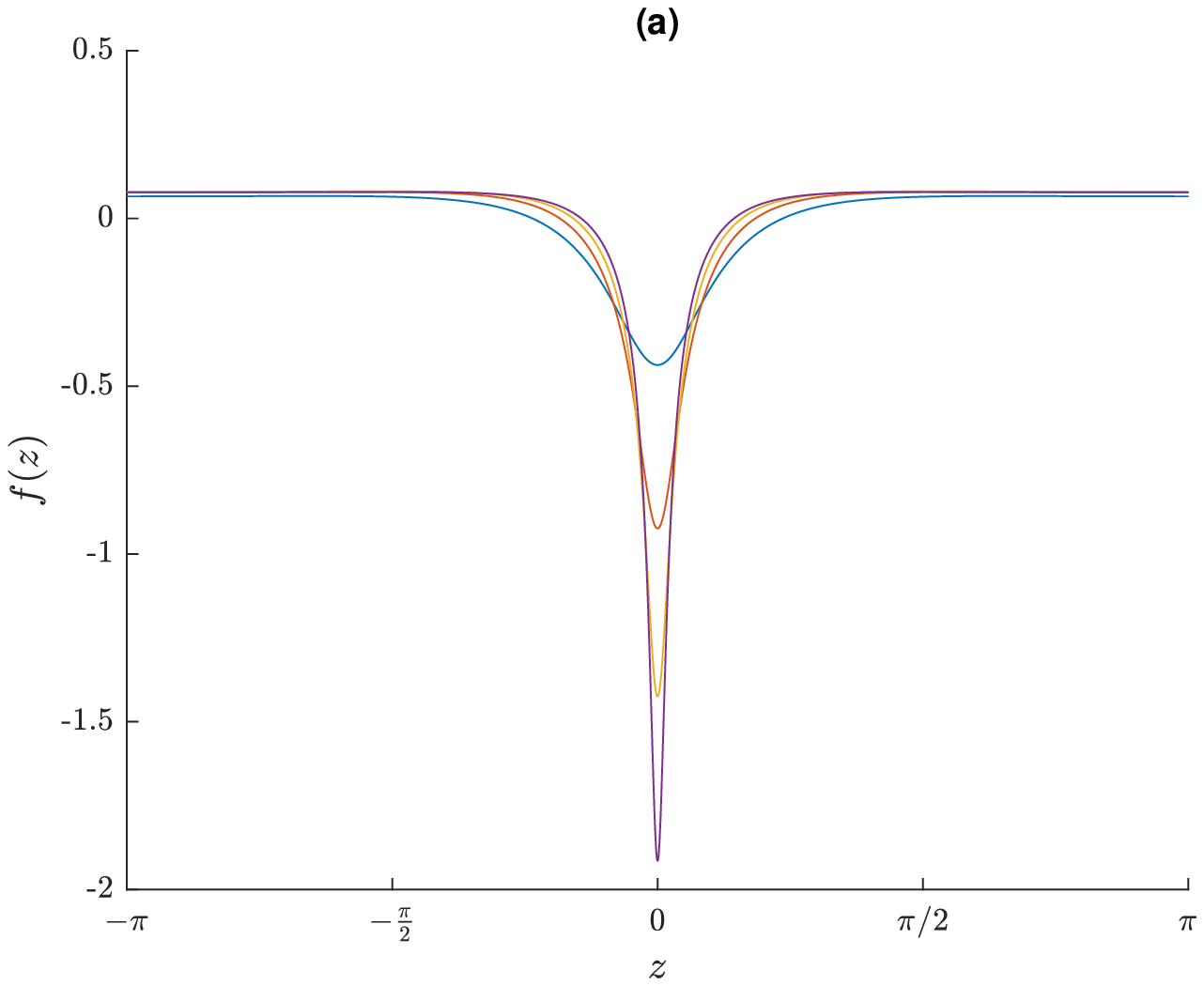}
	\includegraphics[scale=0.4]{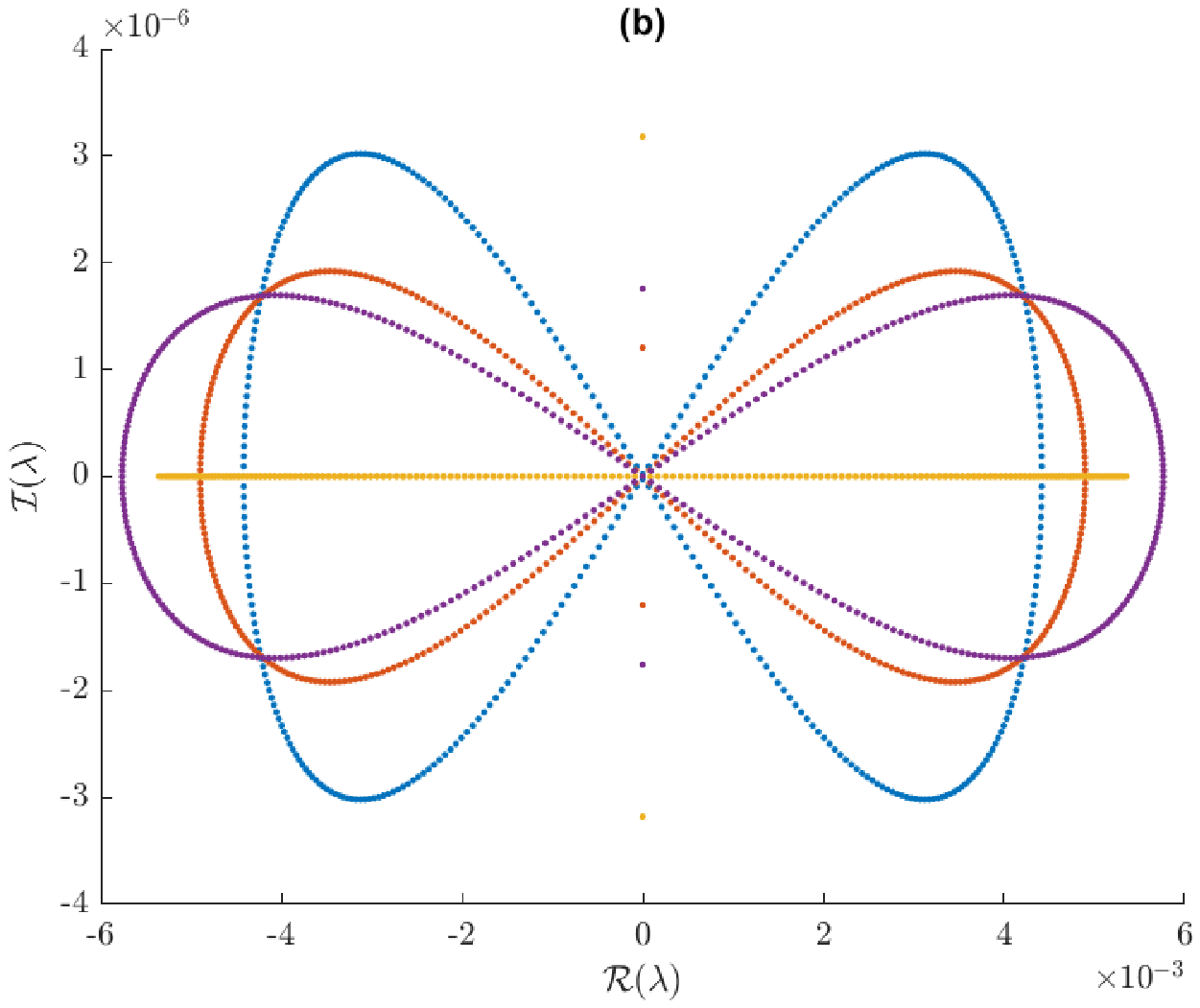}
	\caption{Plots of (a) four representative solutions of the cW equation with $T=\frac{1}{3}$ and $k=1$ and (b) their stability spectra.}
	\label{fig:thirdkOneSolutions}
	\end{center}
\end{figure}

\begin{figure}
	\begin{center}
	\includegraphics[scale=0.4]{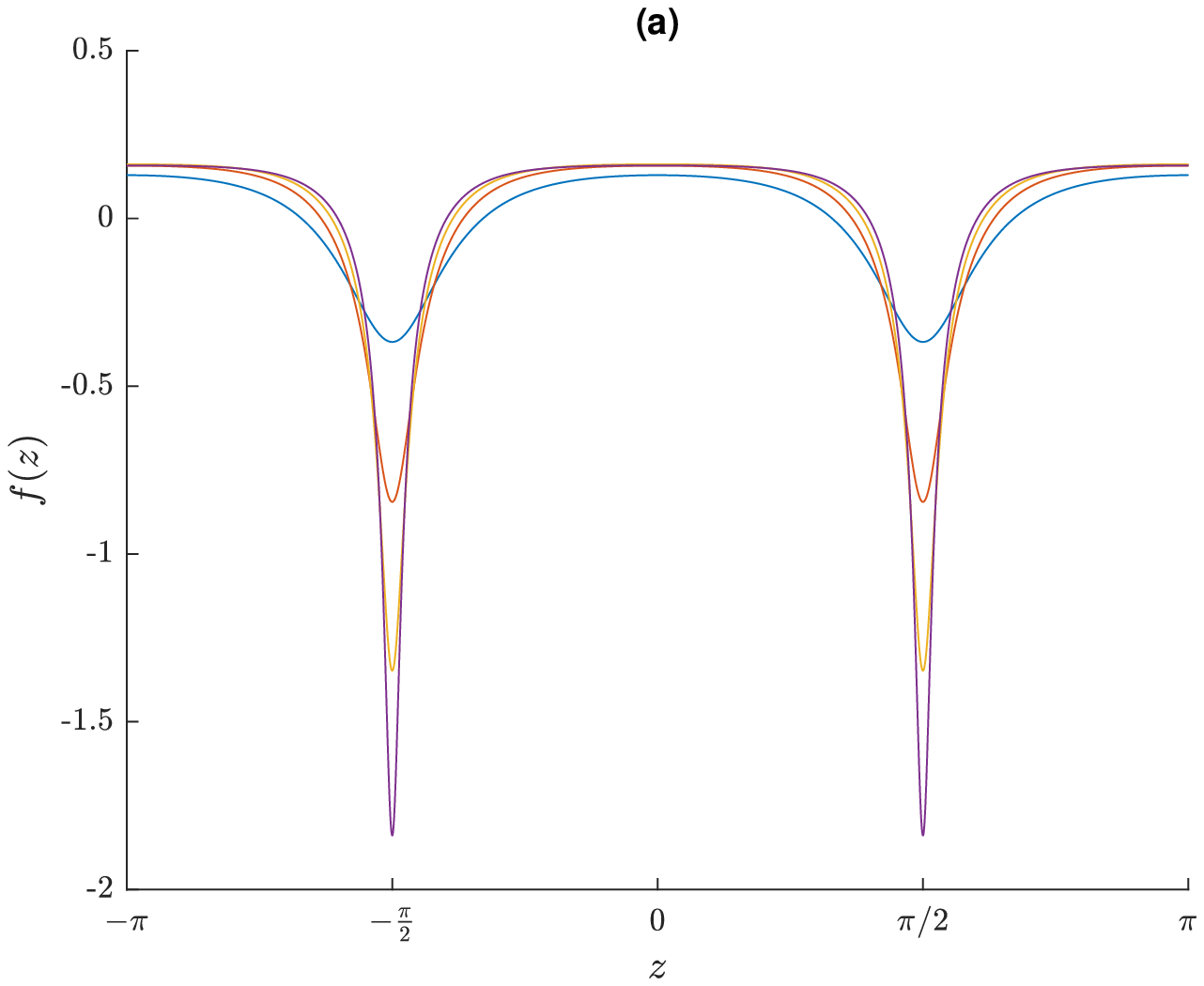}
	\includegraphics[scale=0.4]{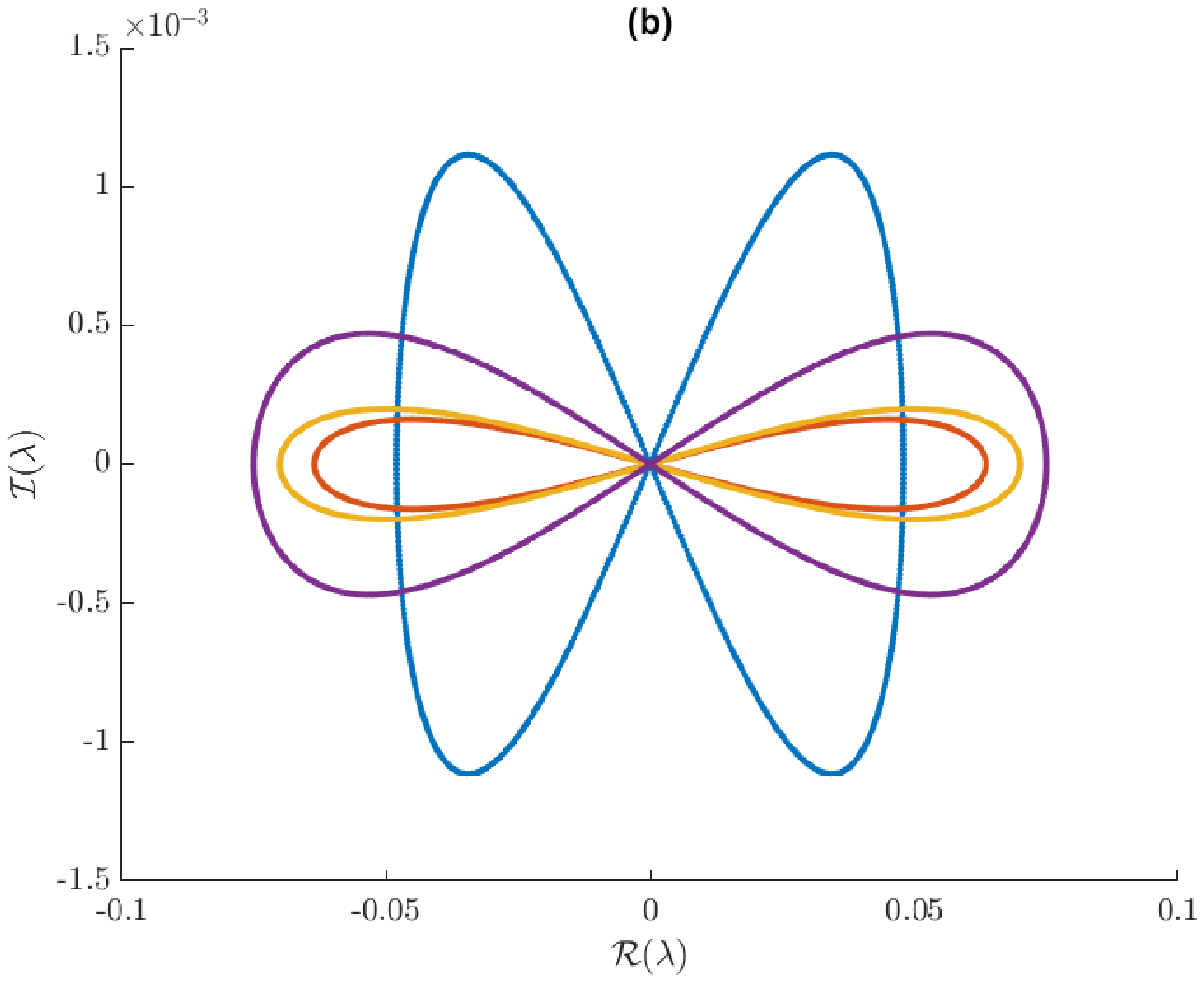}
	\caption{Plots of (a) four representative solutions of the cW equation with $T=\frac{1}{3}$ and $k=2$ and (b) their stability spectra.}
	\label{fig:thirdkTwoSolutions}
	\end{center}
\end{figure}

Figures \ref{fig:thirdkOneSolutions} and \ref{fig:thirdkTwoSolutions} include plots of $k=1$ and $k=2$ solutions and their stability spectra for $T=\frac{1}{3}$.  All eight of these solutions are unstable and their spectra are shaped like horizontal figure 8s.  As wave height increases, the maximal instability growth rate also increases.  The $k=2$ solutions have larger instability growth rates than the corresponding $k=1$ solutions with the same wave height.

\begin{figure}
	\begin{center}
	\includegraphics[scale=0.4]{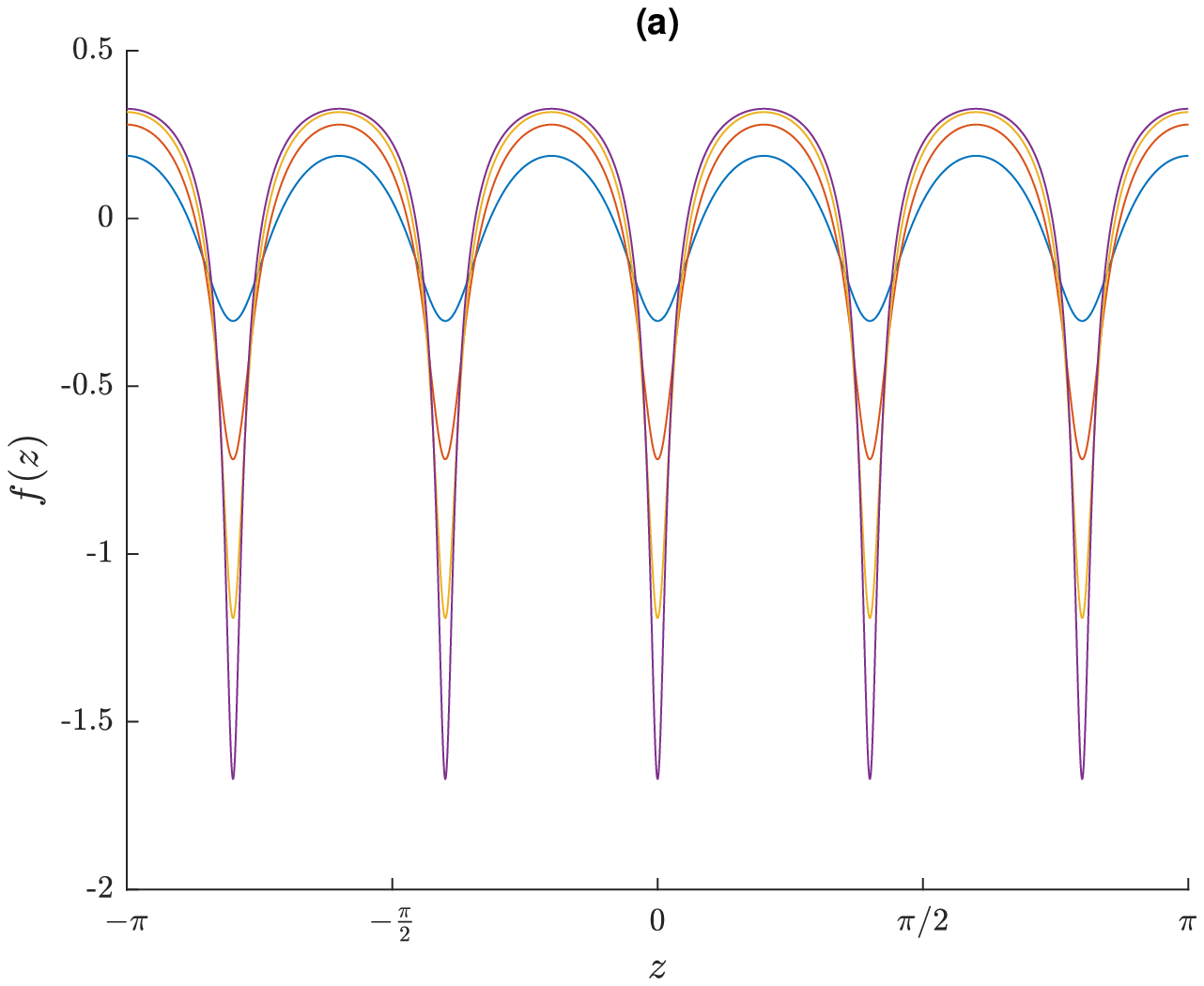}
	\includegraphics[scale=0.4]{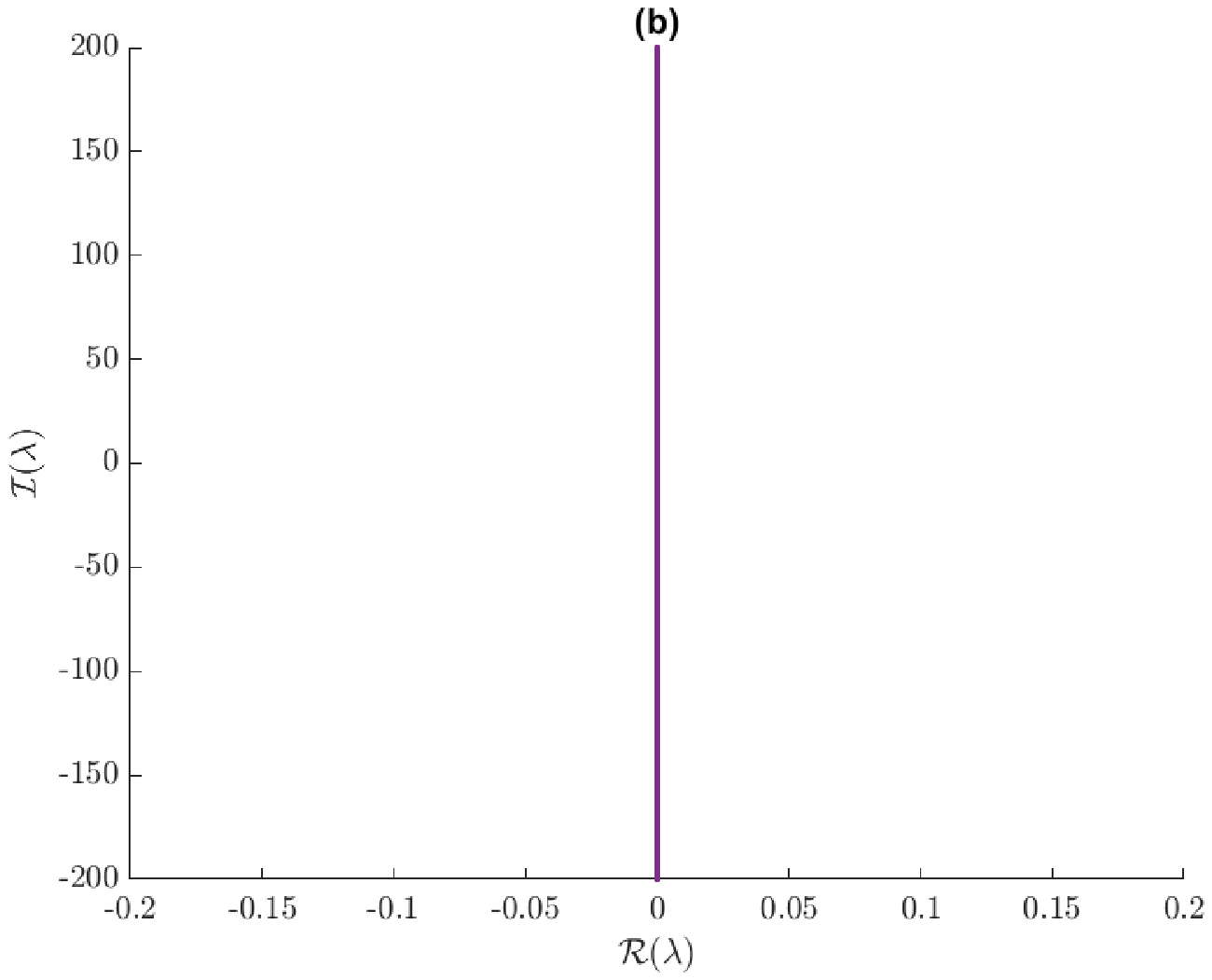}
	\caption{Plots of (a) four representative solutions of the cW equation with $T=\frac{1}{3}$ and $k=5$ and (b) their stability spectra.}
	\label{fig:thirdkFiveSolutions}
	\end{center}
\end{figure}

Figures \ref{fig:thirdkFiveSolutions} includes plots of the $k=5$ solutions and their stability spectra for $T=\frac{1}{3}$.  Other than their periods, these solutions appear to be qualitatively similar to the $k=1$ and $k=2$ solutions.  However, their stability spectra lie completely on the $\mathcal{I}(\lambda)$ axis.  This means that all four of these solutions,regardless of their wave height, are spectrally stable.  This is quite a surprising result.  This suggests that, for this value of $T$, there are bands and gaps in $k$ space where periodic, traveling-wave solutions to the cW equation are stable/unstable.

\subsubsection{Surface tension parameter $T=0.4$}

\begin{figure}
	\begin{center}
	\includegraphics[width=12cm]{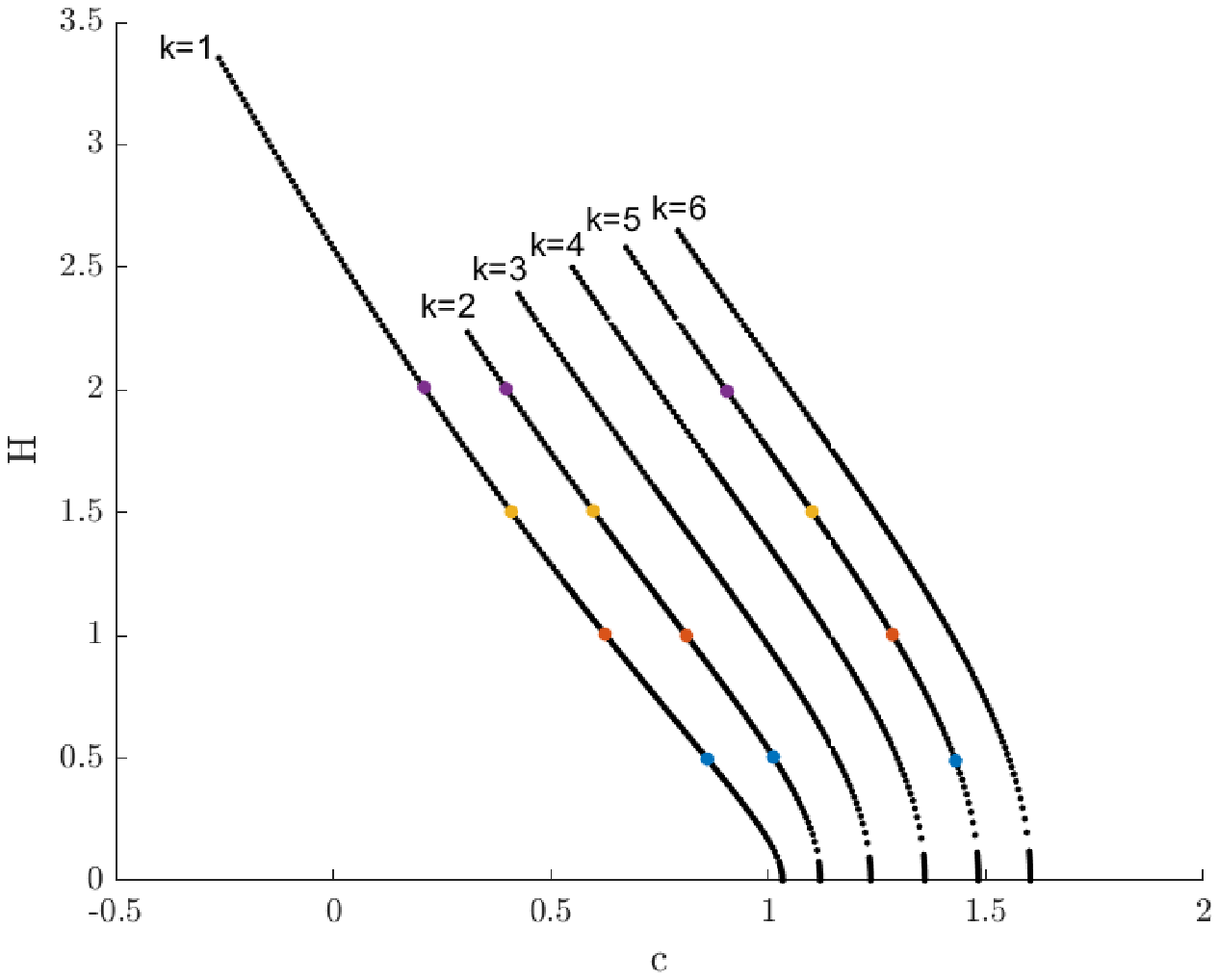}
	\caption{A portion of the bifurcation diagram for the cW equation with $T=0.4$.  The colored dots correspond to solutions that are examined in more detail below.}
	\label{fig:fourBifurcation}
	\end{center}
\end{figure}

Figure \ref{fig:fourBifurcation} includes a portion of the bifurcation diagram for $T=0.4$.  The colored dots correspond to solutions that are examined in more detail below.  These solutions have approximately the same wave heights as the colored solutions examined in other sections.  The bifurcation diagram shows that as wave speed decreases, wave height increases for all branches (that we examined).

\begin{figure}
	\begin{center}
	\includegraphics[scale=0.4]{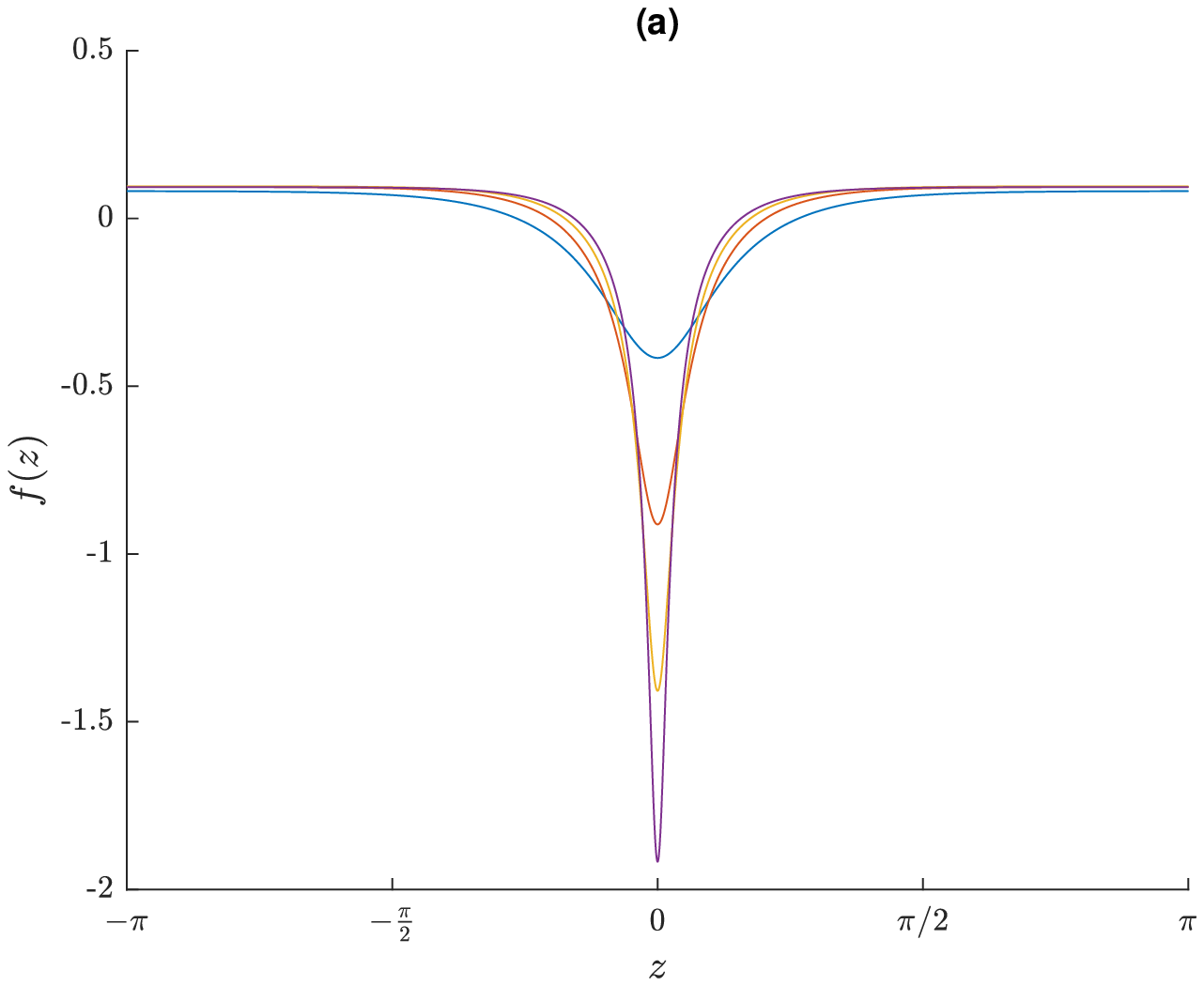}
	\includegraphics[scale=0.4]{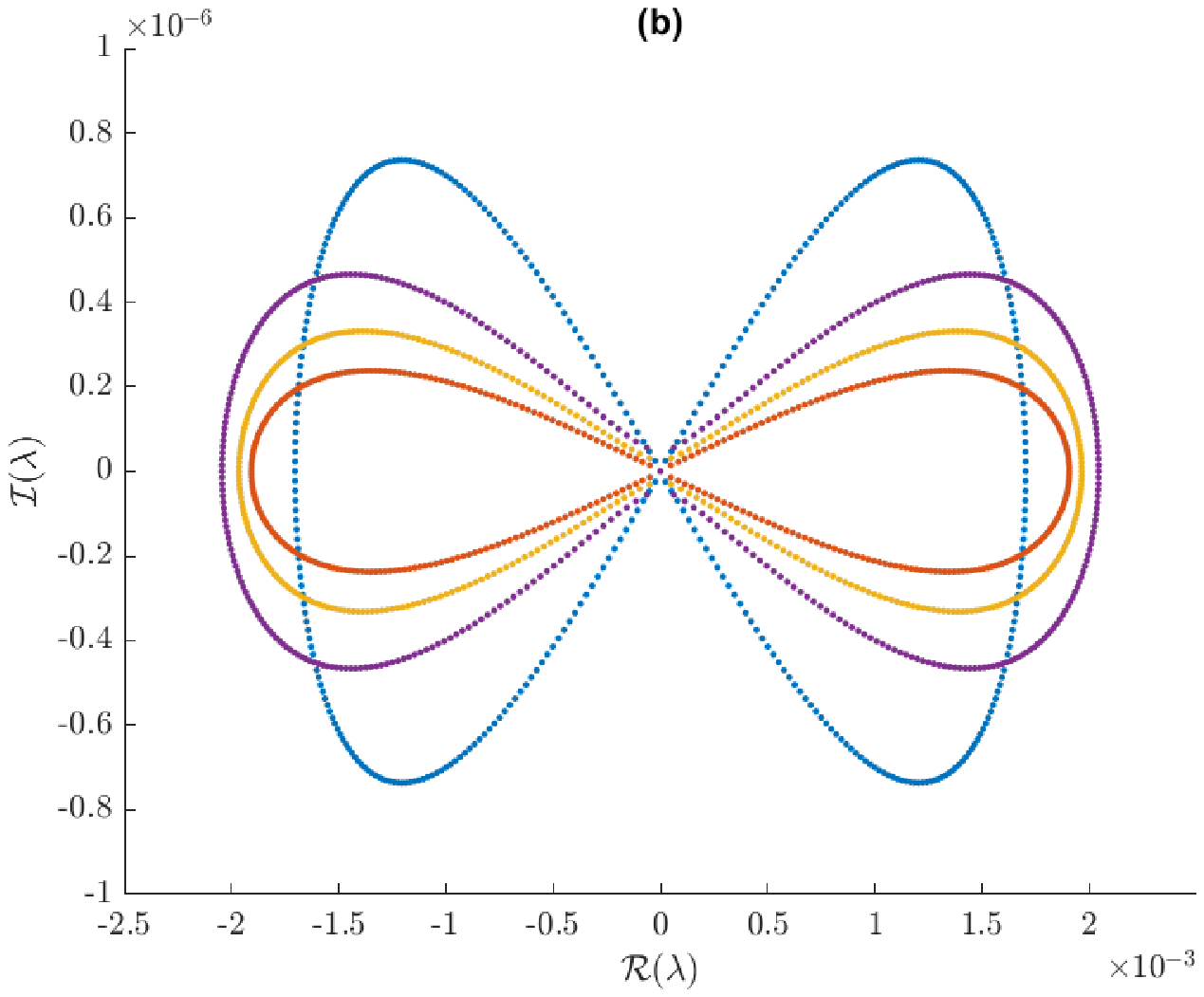}
	\caption{Plots of (a) four representative solutions of the cW equation with $T=0.4$ and $k=1$ and (b) their stability spectra.}
	\label{fig:fourkOneSolutions}
	\end{center}
\end{figure}

\begin{figure}
	\begin{center}
	\includegraphics[scale=0.4]{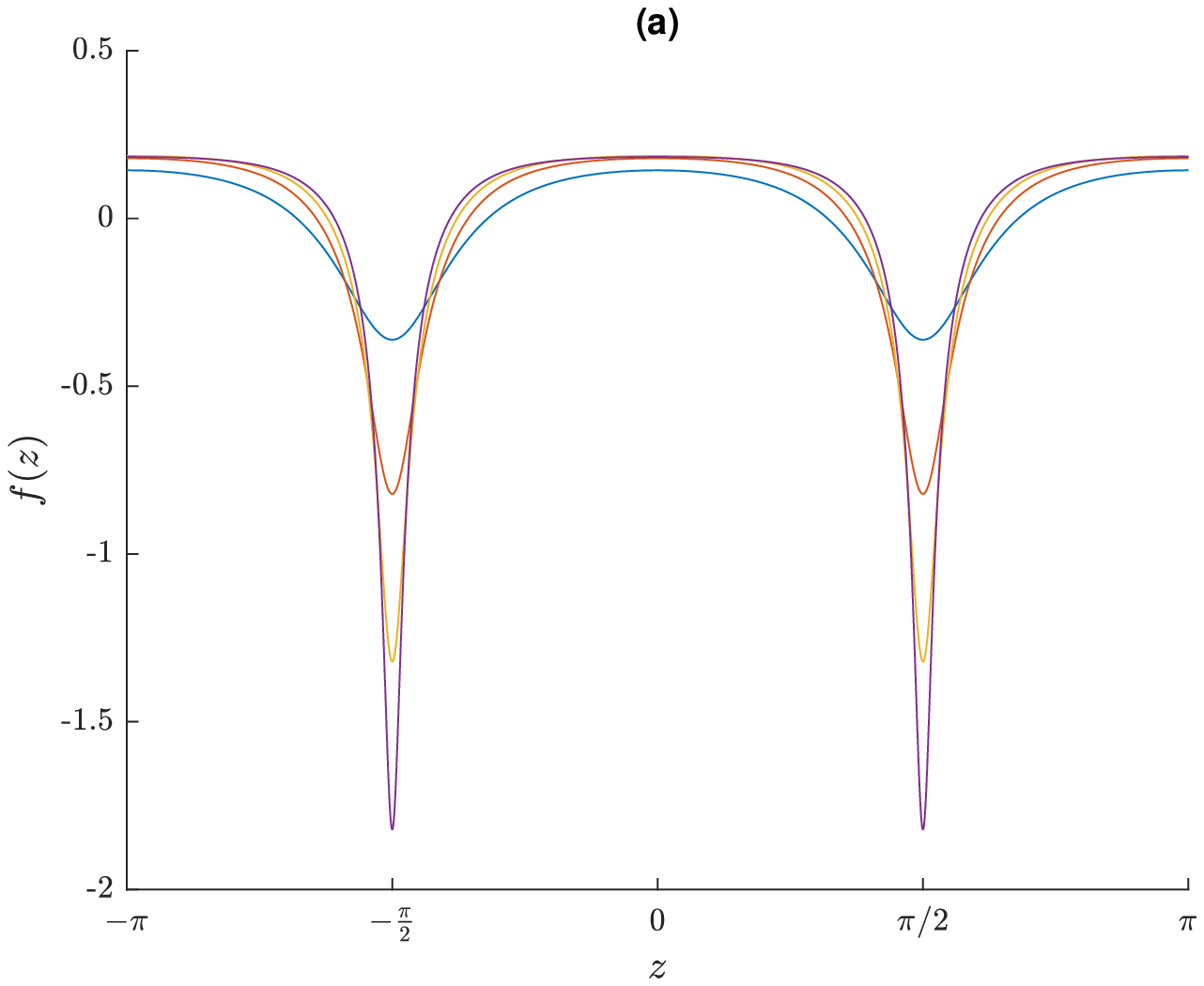}
	\includegraphics[scale=0.4]{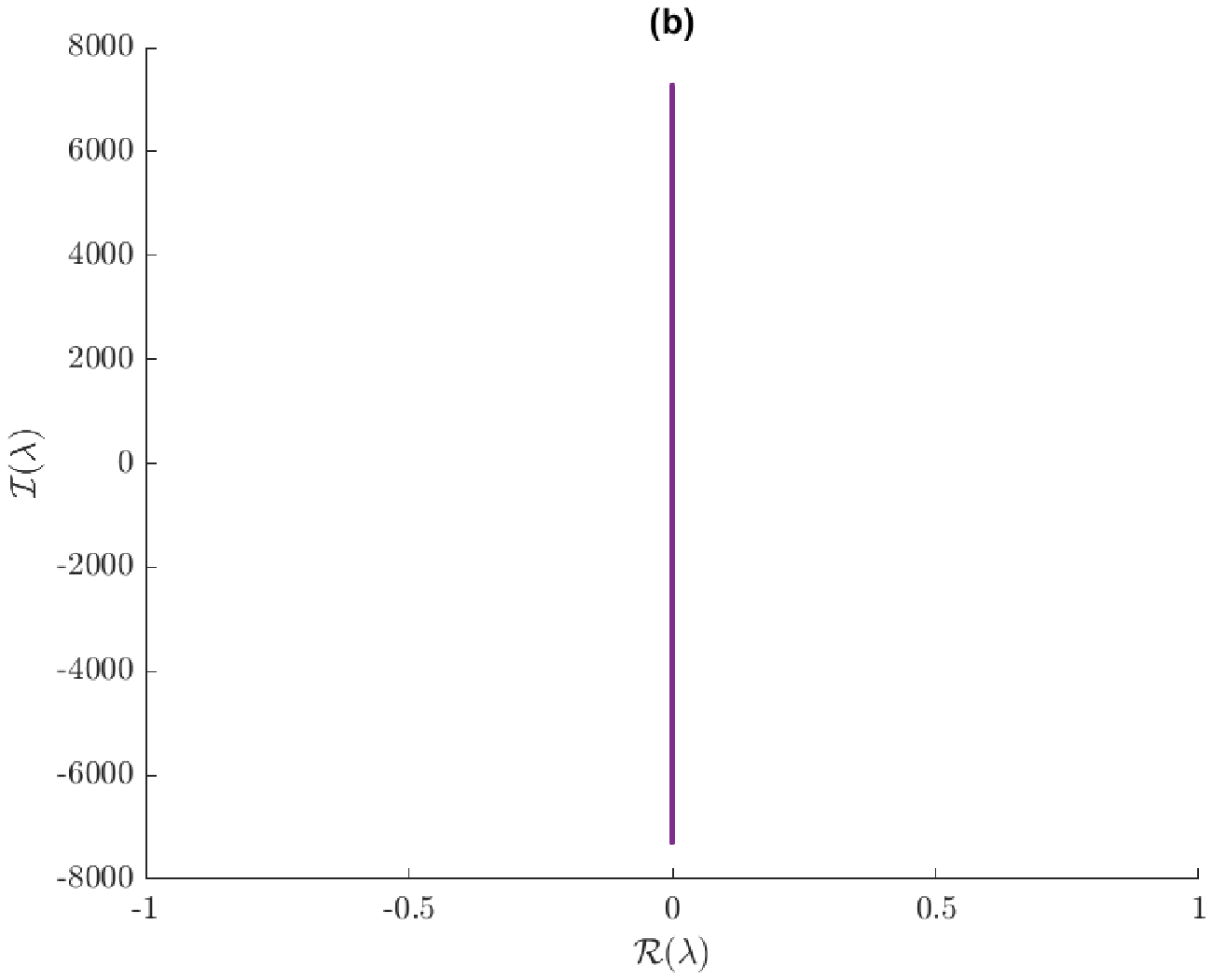}
	\caption{Plots of (a) four representative solutions of the cW equation with $T=0.4$ and $k=2$ and (b) their stability spectra.}
	\label{fig:fourkTwoSolutions}
	\end{center}
\end{figure}

\begin{figure}
	\begin{center}
	\includegraphics[scale=0.4]{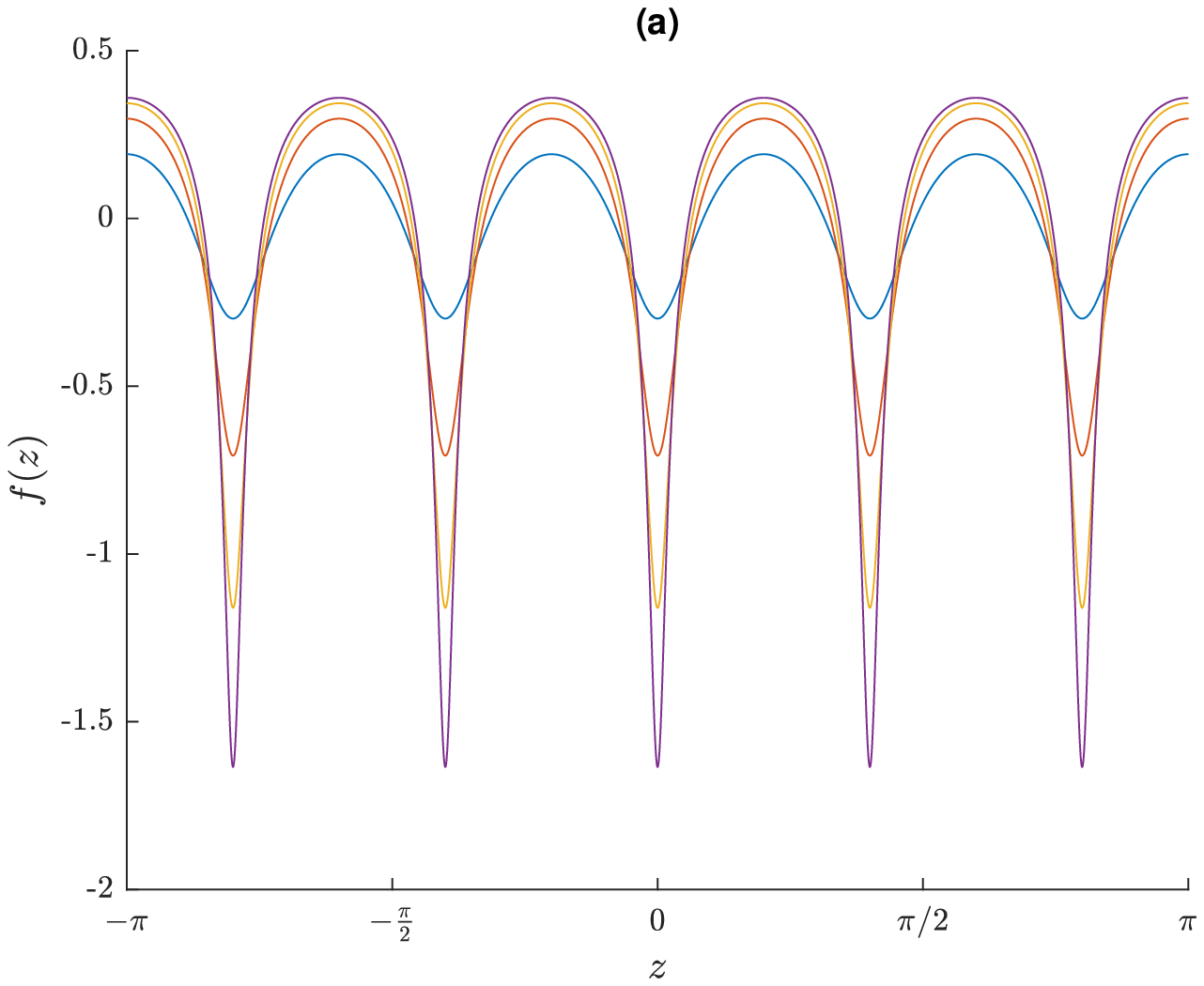}
	\includegraphics[scale=0.4]{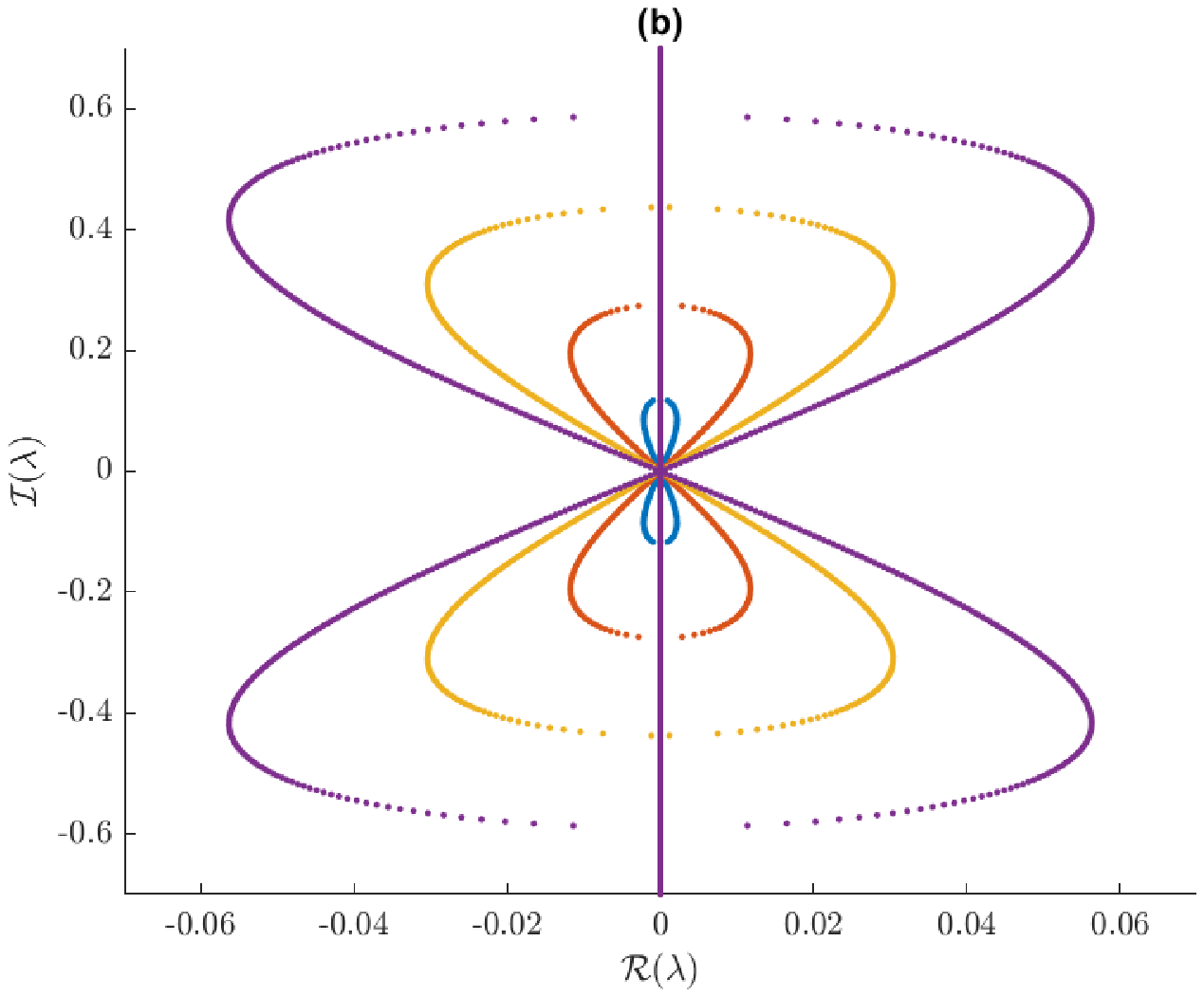}
	\caption{Plots of (a) four representative solutions of the cW equation with $T=0.4$ and $k=5$ and (b) their stability spectra.}
	\label{fig:fourkFiveSolutions}
	\end{center}
\end{figure}

Figure \ref{fig:fourkOneSolutions} includes plots of four $k=1$ solutions to the cW equation with $T=0.4$ and their stability spectra.  All four of these solutions are unstable and the growth rate of the instabilities increases as the wave height of the solution increases.  Figure \ref{fig:fourkTwoSolutions} shows that all four $k=2$ solutions are stable while Figure \ref{fig:fourkFiveSolutions} shows that all four $k=5$ solutions are unstable.  This suggests that, for this value of $T$, there are bands and gaps in $k$ space where periodic, traveling-wave solutions to the cW equation are stable/unstable.  These bands and gaps are likely related to the bands and gaps in the $T=\frac{1}{3}$ case, but we were not able to find a simple relationship between them.

\subsubsection{Surface tension parameter $T\approx0.1582$}
\label{specialT}

\begin{figure}
\begin{center}
\includegraphics[scale = 0.6]{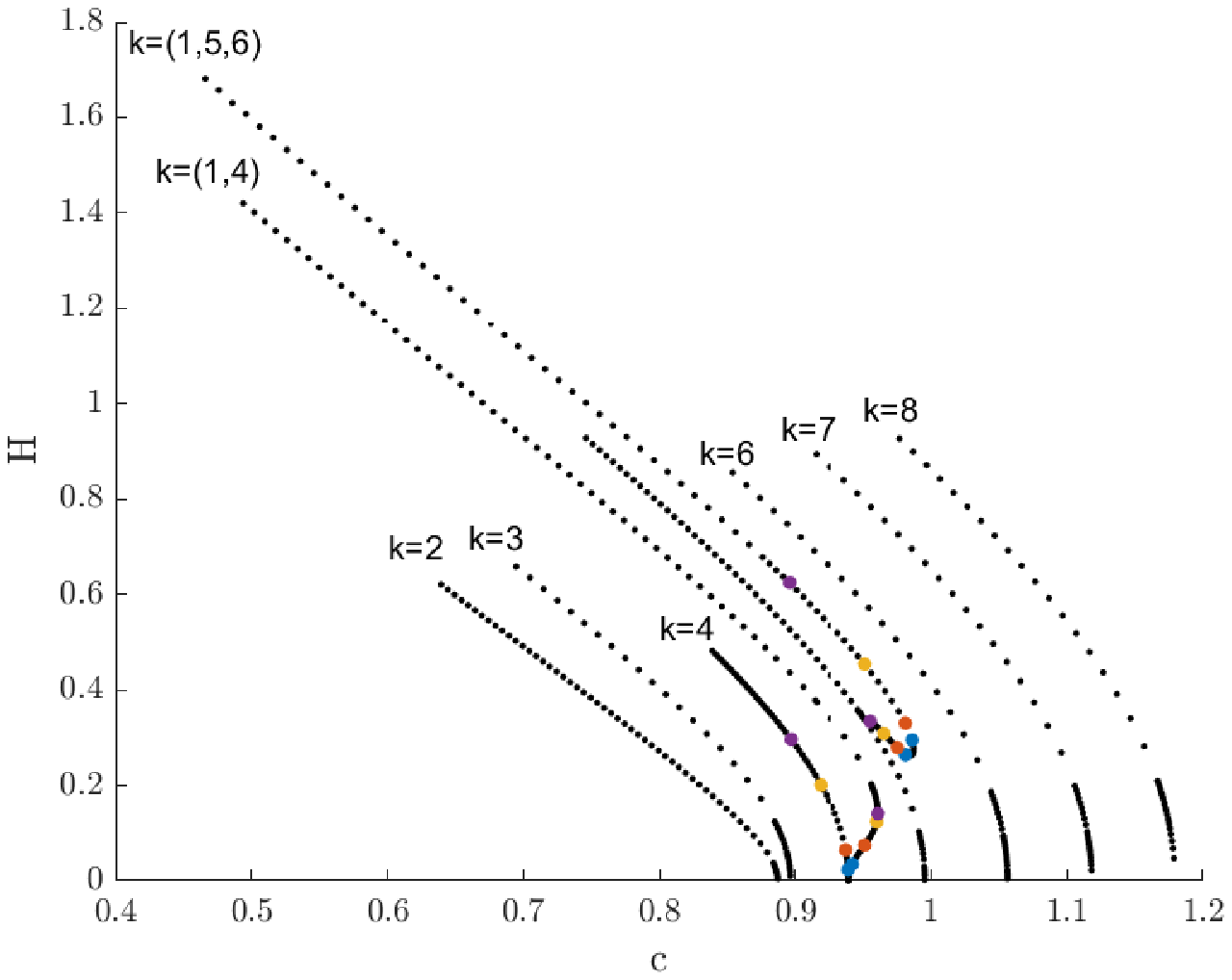}
\caption{A portion of the bifurcation diagram for the cW equation with $T\approx 0.1582$.  The colored dots correspond to solutions that are examined in more detail below.}
\label{fig:onefourBifurcation}
\end{center}
\end{figure}

Remonato \& Kalisch~\cite{RemonatoKalisch} presented the following formula which allows $T$ values to be chosen so that solutions corresponding to any two $k$ values will have the same wave speed in the small-amplitude limit
\begin{equation}
	T=T (k_{1}, k_{2}) = \frac{k_{1}\tanh(k_{2})-k_{2}\tanh(k_{1})}{k_{1}k_{2}(k_{1}\tanh(k_{1})-k_{2}\tanh(k_{2}))}.
\end{equation}
Using this formula with $k_1=1$ and $k_2=4$ gives $T\approx0.1582$, which is the final $T$ value we examine.  A portion of the corresponding bifurcation diagram is included in Figure \ref{fig:onefourBifurcation}.  The fact that the $k=1$ and $k=4$ solutions have the same speed in the small-amplitude limit is exemplified by the fact that there are two branches leaving the same point on the $c$ axis near $c=0.94$.  These branches correspond to the $k=4$ solution and a $k=(1,4)$ branch.  Here the notation $k=(a,b)$ means that the solution is composed of a linear combination of the $k=a$ and $k=b$ wavenumbers in the small-amplitude limit.  We were unable to isolate the $k=1$ solution.  This may be related to the fact that $1<k^*\approx2.29$ when $T\approx0.1582$.  Similarly, we were not able to compute solutions on the $k=2$ branch in this case.

\begin{figure}
\begin{center}
  \includegraphics[scale=0.4]{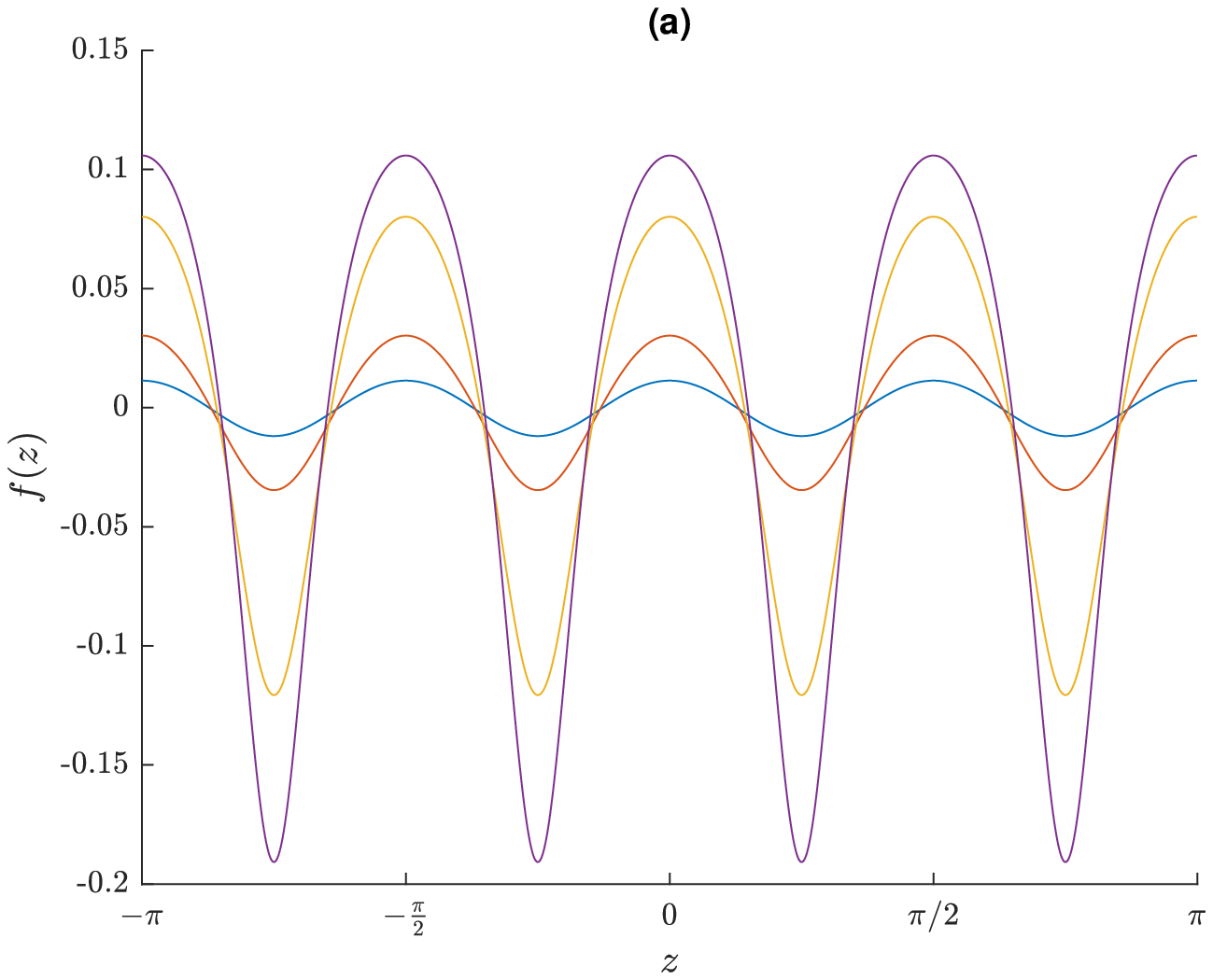}
  \includegraphics[scale=0.4]{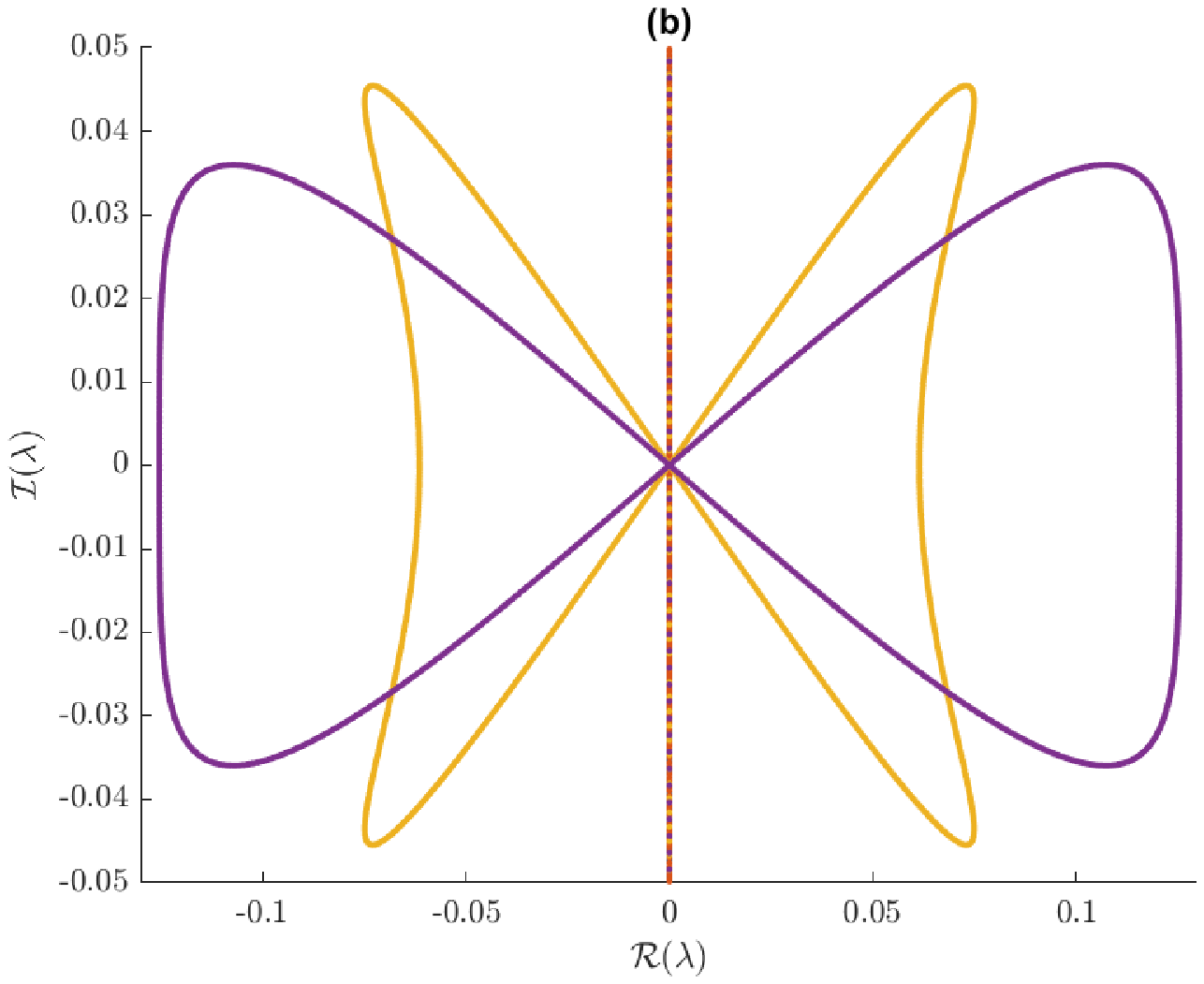}
\caption{Plots of (a) four representative solutions of the cW equation with $T\approx0.1582$ and $k=4$ and (b) their stability spectra.}
\label{fig:onefourkFourSolutions}
\end{center}
\end{figure}

Figure \ref{fig:onefourkFourSolutions} shows that solutions on the $k=4$ branch with small wave height are stable, while those with large wave height are unstable.  These solutions do not have the striking delta-function-like, wave of depression form that the cW solutions presented above have.  As wave height increases, the growth rates of the instabilities also increase.

\begin{figure}
\begin{center}
  \includegraphics[scale=0.4]{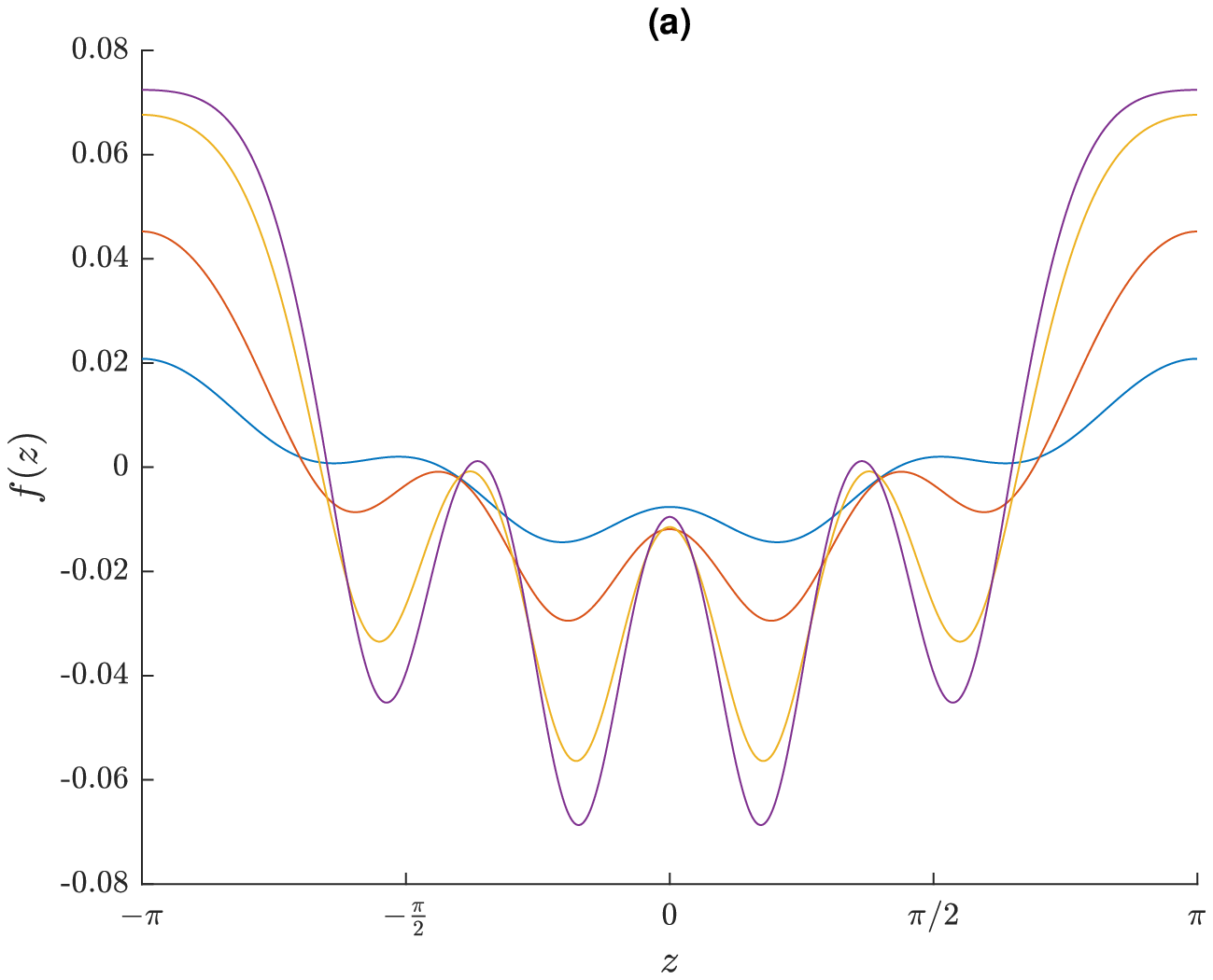}
  \includegraphics[scale=0.4]{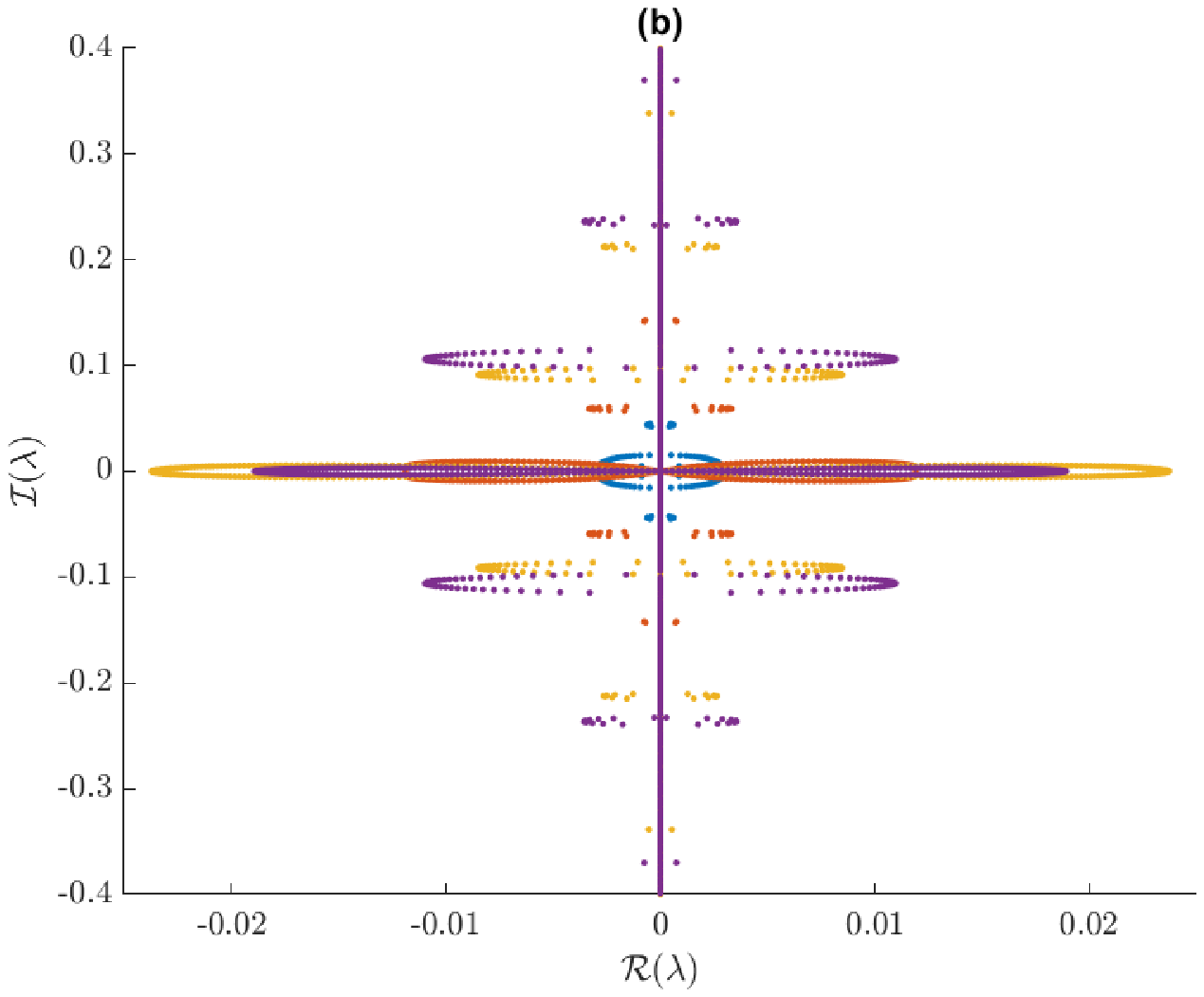}
\caption{Plots of (a) four representative solutions of the cW equation with $T\approx0.1582$ and $k=(1,4)$ and (b) their stability spectra.}
\label{fig:onefourkOneFourSolutions}
\end{center}
\end{figure}

Figure \ref{fig:onefourkOneFourSolutions} includes plots of four representative $k=(1,4)$ solutions and their stability spectra.  These solutions do not have the delta-function-like form of the majority of the other cW solutions.  The spectra of these $k=(1,4)$ solutions are similar to those in Figure \ref{fig:twokOneSolutions}(b).  This is likely due to the fact that both of these sets of solutions have more than one dominant Fourier mode.  Each spectrum has a horizontal figure 8 centered at the origin and six bubbles centered on the $\mathcal{I}(\lambda)$ axis.  The solution with the largest wave height is not the most unstable solution.

\begin{figure}
\begin{center}
  \includegraphics[scale=0.4]{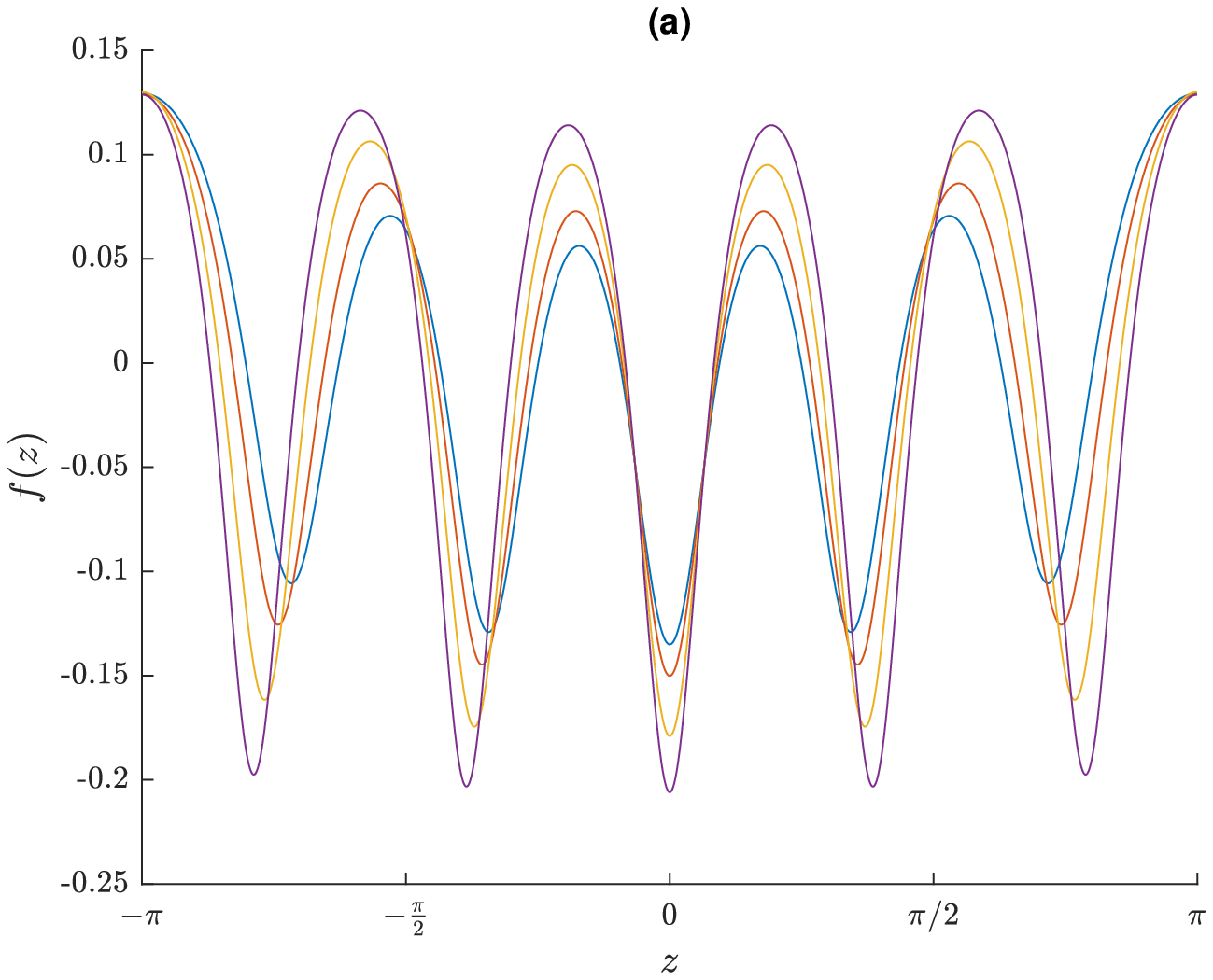}
  \includegraphics[scale=0.4]{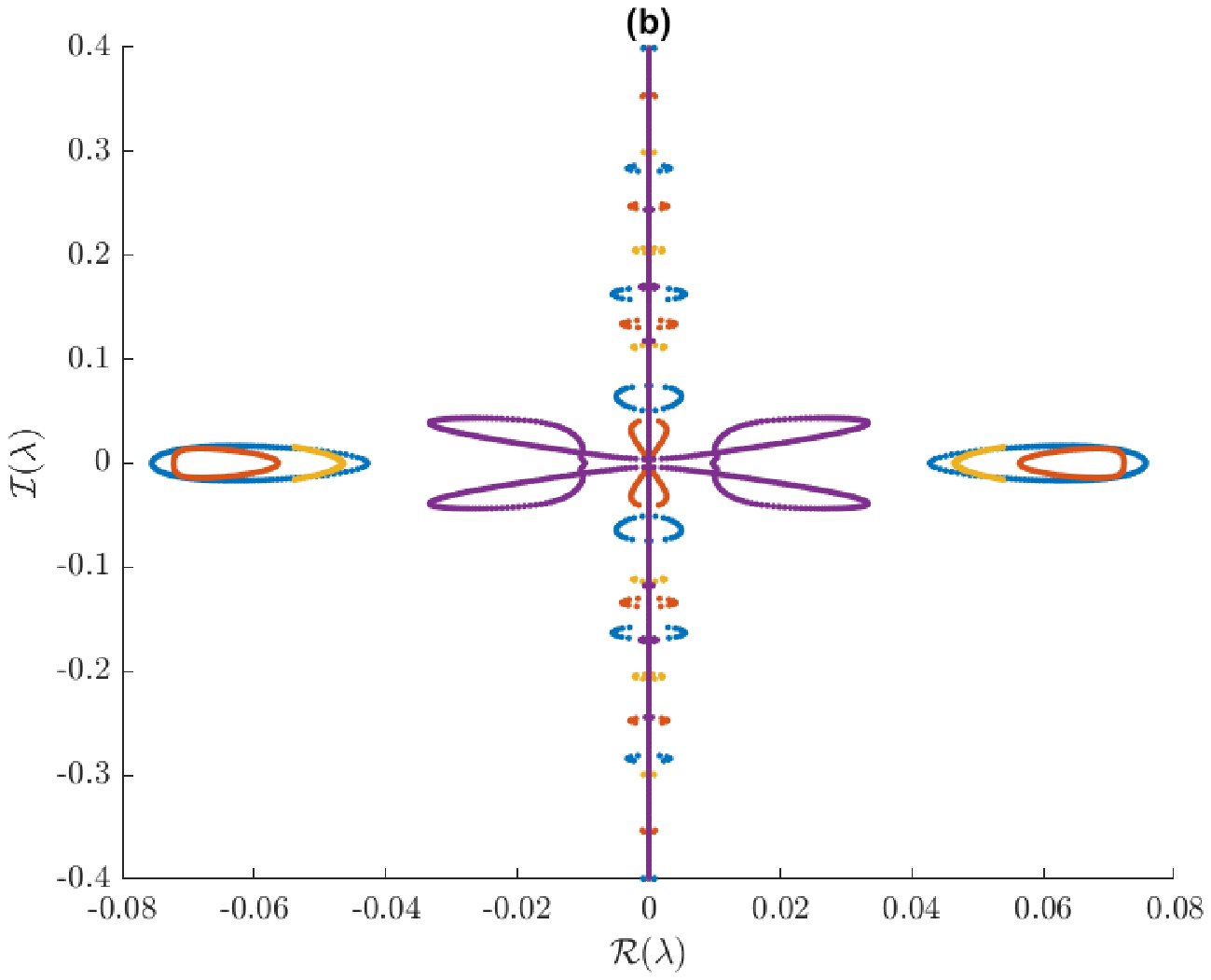}
  \caption{Plots of (a) four representative solutions of the cW equation with $T\approx0.1582$ and $k=(1,5)$ and (b) their stability spectra.}
\label{fig:onefourkOneFiveSolutions}
\end{center}
\end{figure}

\begin{figure}
\begin{center}
  \includegraphics[scale=0.4]{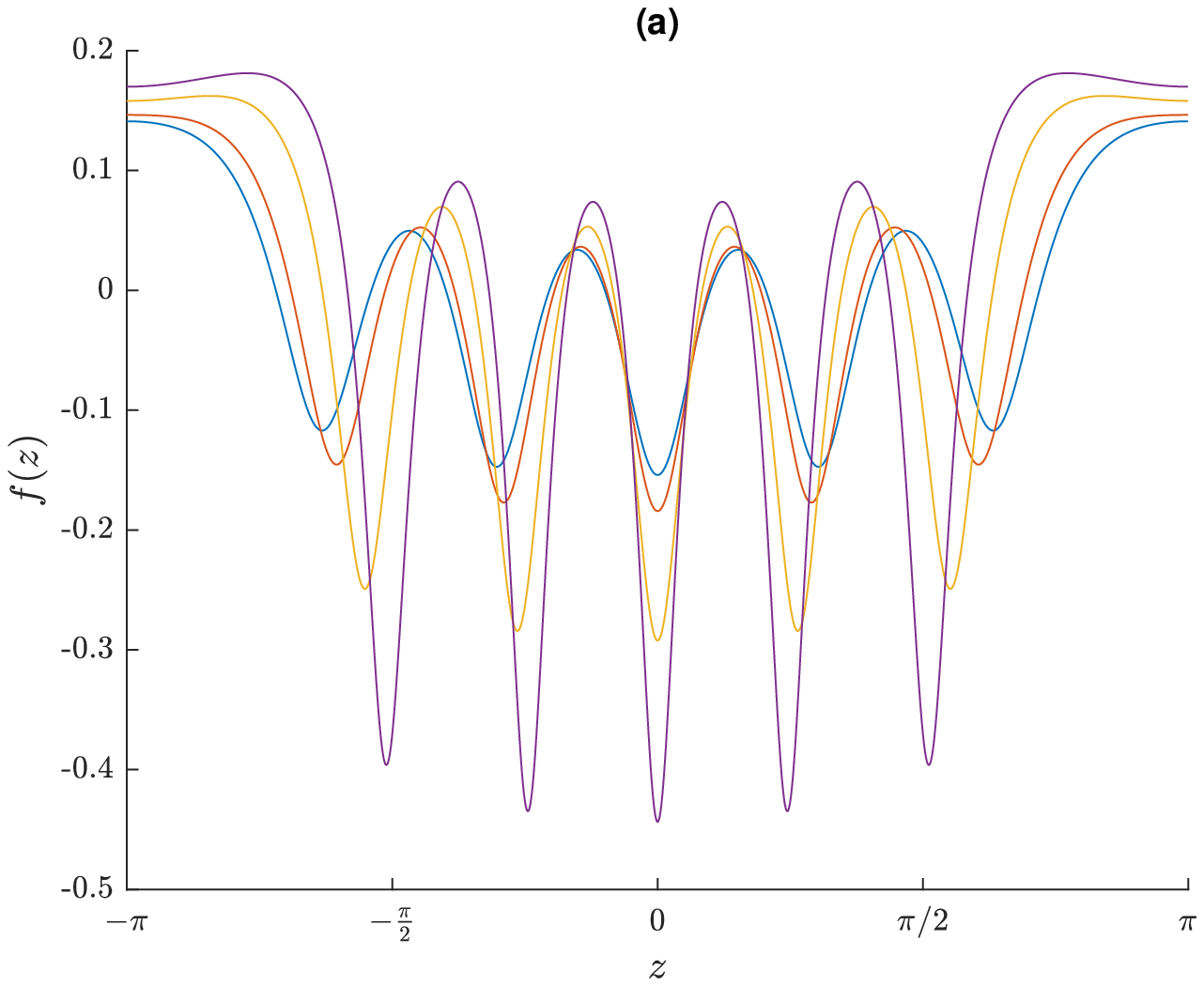}
  \includegraphics[scale=0.4]{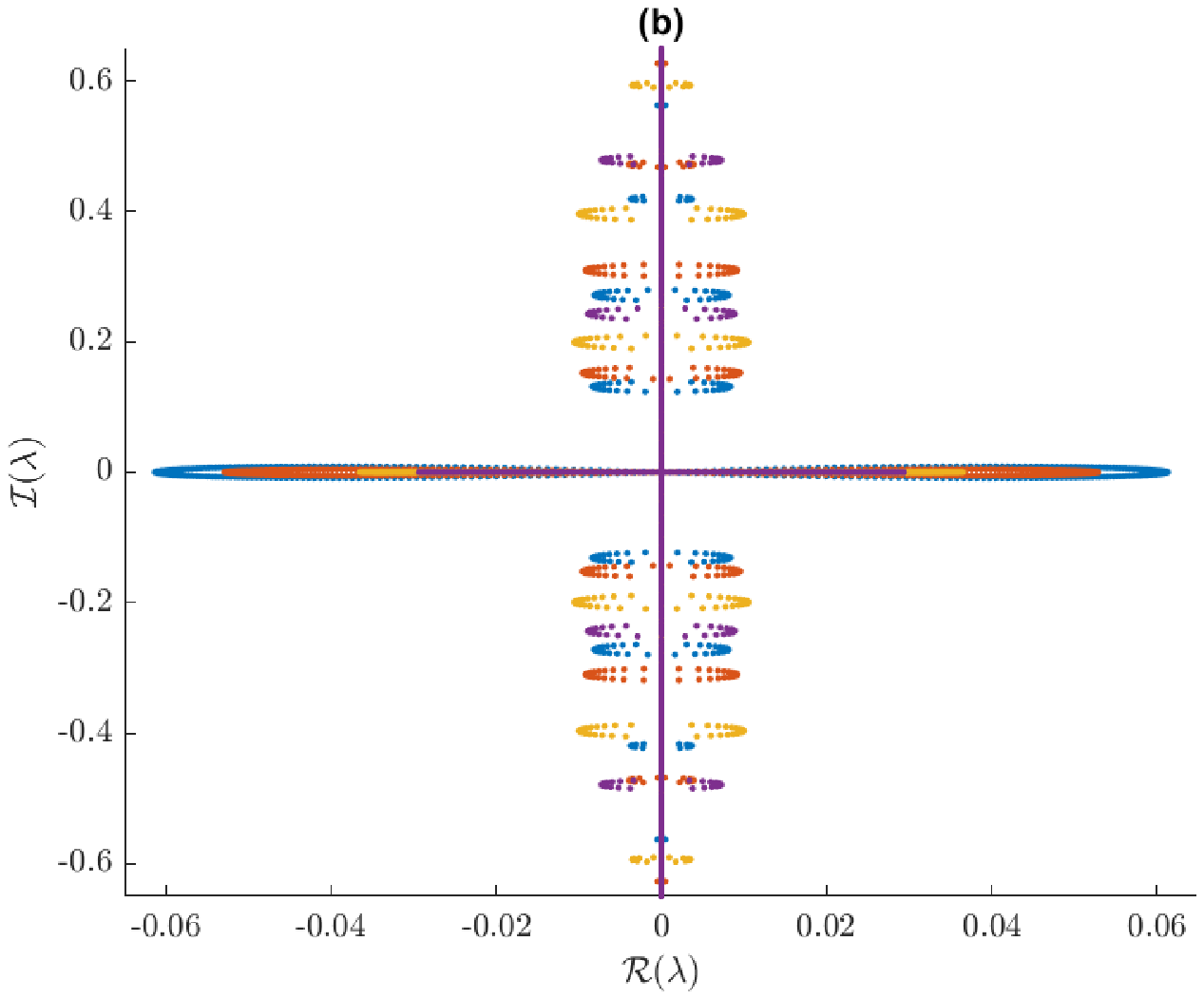}
  \caption{Plots of (a) four representative solutions of the cW equation with $T\approx0.1582$ and $k=(1,5,6)$ and (b) their stability spectra.}
\label{fig:onefourkOneFiveSixSolutions}
\end{center}
\end{figure}

Figure \ref{fig:onefourBifurcation} shows that there is a secondary branch that splits off from the $k=5$ branch at $H\approx0.39$.  As expected, the solutions along this branch do not have a single dominant wavenumber.  The solutions on this branch are $k=(1,5)$ solutions until the branch curves around and heads upward.  The solutions after this turning point are $k=(1,5,6)$ solutions.  Figure \ref{fig:onefourkOneFiveSolutions}(a) includes plots of four $k=(1,5)$ solutions and their stability spectra.  All four of these solutions are unstable and have complicated spectra.  Figure \ref{fig:onefourkOneFiveSixSolutions} includes plots of four representative $k=(1,5,6)$ solutions and shows that they are unstable.  All four solutions have horizontal figure 8s centered at the origin and four bubbles centered along the $\mathcal{I}(\lambda)$ axis.  The solution with smallest wave height is the most unstable.

\section{Summary}
\label{Summary}

Bottman \& Deconinck~\cite{BD} proved that all traveling-wave solutions of the KdV equation are stable.  This is quite different than the Whitham equation where all large-amplitude solutions are unstable~\cite{WhithamStability} and only small-amplitude solutions with a wavenumber larger than $k=1.145$ are unstable~\cite{MatVera,WhithamStability}.  

We began by examining large-amplitude, periodic, traveling-wave solutions to the Whitham equation (zero surface tension).  We found that all such solutions are unstable and that their stability spectra undergo two bifurcations as wave height increases.  

Next, we examined periodic, traveling-wave solutions to the capillary-Whitham equation with four different $T$ values.  As expected, we found that the cW solutions and their stability were more diverse than in the Whitham equation case.  Most of the solutions we examined were waves of depression.  In contrast, all periodic, traveling-wave solutions to the Whitham equation are waves of elevation.  We found that as wave height increases, wave speed decreases (and can become negative).  We were not able to determine if the cW equation admits a solution with maximal wave height.  In contrast, the Whitham equation has a solution with maximal wave height.  We computed periodic, traveling-wave solutions with wavenumbers $k>k^*$ where $k^*$ is the location of the local minimum of the Fourier multiplier $\mathcal{K}$.  We were not able to compute any periodic, traveling-wave solutions with $k<k^*$.  In addition to computing a variety of solutions with a single dominant wavenumber, we computed four families of solutions that had multiple dominant wavenumbers.

We examined the stability of all of the cW solutions we computed.  We found that some were stable and others were unstable.  There appear to be bands and gaps in $k$ and $T$ space that separate small- and large-amplitude solutions to the cW equation by stability.  The exact structure of these bands and gaps remains an open question.  If the solutions had a single dominant wavenumber, then the maximal instability growth rate increased with wave height.  If the solutions had multiple dominant wavenumbers, there was not a simple relationship between instability growth rates and wave heights.  We found regions of $k$ and $T$ space where all solutions appeared to be stable, regardless of wave height.  Finally, we found some solutions for which the modulational (Benjamin-Feir) instability was the dominant instability and other solutions that had other dominant instabilities.

We thank Mats Ehrnstr\"om, Vera Hur, Mat Johnson, and Logan Knapp for helpful discussions.  This material is based upon work supported by the National Science Foundation under grant DMS-1716120.

\end{document}